\documentclass[fleqn,usenatbib]{mnras}

\usepackage{newtxtext,newtxmath}
\usepackage[T1]{fontenc}

\DeclareRobustCommand{\VAN}[3]{#2}
\let\VANthebibliography\thebibliography
\def\thebibliography{\DeclareRobustCommand{\VAN}[3]{##3}\VANthebibliography}

\usepackage{graphicx}	% Including figure files
\usepackage{amsmath}	% Advanced maths commands
\usepackage{siunitx}
\title[Spiral arm impact on the stellar kinematics]{Impacts of the Local arm on the local circular velocity inferred from the Gaia DR3 young stars in the Milky Way}

\author[]{
Aisha S. Almannaei,$^{1}$\thanks{E-mail: aisha.almannaei.18@ucl.ac.uk (KTS)}
Daisuke Kawata,$^{1}$
Junichi Baba $^{2,3}$
Jason A. S. Hunt$^{4}$
George Seabroke$^{1}$
Ziyang Yan$^{5}$
\\
$^{1}$Mullard Space Science Laboratory, University College London, Holmbury St Mary, Dorking, Surrey RH5 6NT, UK\\
$^{2}$Amanogawa Galaxy Astronomy Research Center, Kagoshima University, 1–21–35 Korimoto, Kagoshima 890-0065, Japan\\
$^{3}$National Astronomical Observatory of Japan, Mitaka-shi, Tokyo 181-8588, Japan\\
$^{4}$School of Mathematics and Physics, University of Surrey, Guildford, GU2 7XH, UK.\\
$^{5}$Department of Physics and Astronomy, University College London, Gower Street, London WC1E 6BT, UK\\
}

\date{Accepted XXX. Received YYY; in original form ZZZ}

\pubyear{2015}

\begin{document}
\label{firstpage}
\pagerange{\pageref{firstpage}--\pageref{lastpage}}
\maketitle

\begin{abstract}
A simple one-dimensional axisymmetric disc model is applied to the kinematics of OB stars near the Sun obtained from Gaia DR3 catalogue. The model determines the ‘local centrifugal speed’ $V_\mathrm{c}(R_{0})$ - defined as the circular velocity in the Galactocentric rest frame, where the star would move in a near-circular orbit if the potential is axisymmetric with the local potential of the Galaxy.We find that the $V_\mathrm{c}(R_{0})$ values and their gradient vary across the selected region of stars within the solar neighbourhood. By comparing with an N-body/hydrodynamic simulation of a Milky Way-like galaxy, we find that the kinematics of the young stars in the solar neighbourhood is affected by the Local arm, which makes it difficult to measure $V_\mathrm{c}(R_{0})$. However, from the resemblance between the observational data and the simulation, we suggest that the known rotational velocity gap between the Coma Bernices and Hyades-Pleiades moving groups could be driven by the co-rotation resonance of the Local arm, which can be used to infer the azimuthally averaged circular velocity. We find that $V_\mathrm{c}(R)$ obtained from the $\mathrm{D}<2$ \unit{kpc} sample is well matched with this gap at the position of the Local arm. Hence, we argue that our results from the $\mathrm{D}<2$ kpc sample, $V_\mathrm{c}(R_{0})= 233.95\pm2.24$ \unit{km.s^{-1}}, 
is close to the azimuthally averaged circular velocity rather than the local centrifugal speed, which is influenced by the presence of the Local arm.
\end{abstract}


%%%%%%%%%%%%%%%%%%%%%%%%%%%%%%%%%%%%%%%%%%%%%%%%%%

%%%%%%%%%%%%%%%%% BODY OF PAPER %%%%%%%%%%%%%%%%%%

%%%%%%%%%%%%%%%%%%%%%%
\section{Introduction}
%%%%%%%%%%%%%%%%%%%%%%
The Milky Way’s circular velocity, $V_\mathrm{c}(R)$, and more specifically, its value at the position of the Sun, $V_\mathrm{c}(R_{0})$, provides important insights into the Galaxy’s formation, structure, dynamics and evolution. It is used to infer the local dark matter density, combined with the estimated baryonic mass. The non-uniformity of visible matter within our Galaxy -concentrated mainly within the Galactic disc and bulge- causes variations in the gravitational potential of our Galaxy, which provides significant insights into understanding the dynamics of the stars, gas and dust that we observe. The shape of the $V_\mathrm{c}(R)$ curve also provides constraints on modeling the Galactic disc and its evolution. Moreover, its value is necessary for dark matter direct detection experiments, a missing ingredient to our comprehension of the large-scale structures and matter in the Universe.

However, measuring the value of $V_\mathrm{c}(R_{0})$ is not straightforward, and different measurements yield systematically different results. $V_\mathrm{c}(R_{0})$ is also affected by the presence of spiral arms in the Milky Way, which are regions of high stellar density, dust and star formation. The conventional approach to obtain $V_\mathrm{c}(R_{0})$ is by observing the Sun’s rotation velocity $\Omega_{\odot }$ with respect to the Galactocentric rest frame. This is measured as the proper motion of Sgr A* assuming that Sgr A* is fixed at the center of the Galaxy. By utilizing the Very Long Baseline Interferometry (VLBI), \citet{ReidBrunthaler2020} calculated the proper motion of Sgr A* by observing its apparent motion over time and measuring the change in its position with respect to a background quasar and determined $\Omega_{\odot } = 30.32\pm 0.27$ \unit{km.s^{-1}.kpc^{-1}}. This technique requires an observational measurement of the distance from the Galactic center to the Sun, $R_{0}$, to calculate the Sun’s rotation speed, $V_\mathrm{\phi,\odot} = \Omega_{\odot} R_{0}$. These observations are directly measured, but it is difficult to measure the circular velocity because it is required to know the Sun’s peculiar rotation velocity, $V_{\odot}= V_{\phi,\odot} - V_\mathrm{c}({R_0})$  \citep[][for a review]{BlandHawthorn2016}, which is calculated using the kinematics of the stars in the Galactic disc \citep{Schonrich2010, Bovy2012, Kawata2019}, by using stellar streams \citep{Koposov2010} or by directly measuring its acceleration \citep{Bovy2020}. 

The most common method is using the kinematics of the stars via the axisymmetric model \citep{Bovy2012, Kawata2019, Eilers2019, Nitschai2021}. This model assumes that the Milky Way is axisymmetric around its rotational axis. To measure $V_\mathrm{c}(R_{0})$, young stars such as OB stars and Cepheids are used because their asymmetric drift is minimal, i.e. their mean rotation velocity is estimated to be close to the $V_\mathrm{c}(R_{0})$. Additionally, classical Cepheids are considered to be a good kinematic and structural tracer for this purpose \citep{Kawata2019, BobylevBajkova2023}, because of their accurately measured distances derived from the widely known period-luminosity relationship \citep{Leavitt1908, Leavitt1912}. 

Moreover, spiral arm features, and more specifically, the Local arm, have been previously identified with young stars such as Cepheids, OB stars and massive star forming regions \citep{Xu2018, Reid2019, Xu2021, BobylevBajkova2022}. Young massive stars are associated with the star forming regions which emit strong radio signals and are not affected by interstellar dust. Therefore, the spiral arms feature has been readily traced at a considerable distance of $\sim20$ kpc across the Galactic plane \citep{Sanna2017}. The Sun's position is also closely adjacent to the Local arm, as highlighted by \citet{Reid2019, Poggio2021}. It remains uncertain if the Local arm is a real major arm or a spur that connects the Sagittarius-Carina arm and the Perseus arm \citep{Russeil2007,Xu2016,Xu2018, Miyachi2019}.  If the Local arm is a real massive arm, it should impact the kinematics, in particular, the perturbation velocity of young stars in the Galactic disc stars \citep{Liu2017} or other effects linked to the vertical oscillations in the Galactic disc \citep{Widrow2012,BlandHawthorn2021, Kumar2022, Widmark2022, Yang2023, Asano2023}.

However, the behavior of spiral arms in galaxies is currently a topic of intense debate. There are two leading scenarios for the origin of spiral arms in galaxies. The first scenario is known as the density wave, which serves as a classical view wherein spiral arms are thought of as long-lived and rigidly rotating density wave features \citep{LinShu1964, LinShu1966, BertinLin1996}. The second scenario is known as the transient dynamic spiral-arm scenario, where the spiral arms are short-lived, transient, and recurrent \citep{SellwoodCarlberg1984,Fujii2011,Donghia2013}. The latter scenario has gained more traction in N-body simulations of spiral disc galaxies, where the spiral arms are co-rotating and winding with the stars in every radius \citep{Baba2009, Grand2012a, Grand2012b, Baba2015}.

\citet{Kawata2019} suggested that the non-axisymmetric potential, such as spiral arms, affects the measured $V_\mathrm{c}(R_{0})$ when the axisymmetric disc model is applied especially for the young stars. In other words, the impact of the spiral arm on the kinematics of the disc stars can be observed with the variation of $V_\mathrm{c}(R_{0})$ in different regions of the disc  \citep[][for more details]{Trick2017}. Hence, the derived $V_\mathrm{c}(R_{0})$ is considered as the local centrifugal speed, which is defined as the local circular velocity, where the star would move in a near-circular orbit if the potential is axisymmetric with the local potential. This can be different from the azimuthally averaged circular velocity, $\overline{V_\mathrm{c}(R_0)}$, which better represents the total mass of the Galaxy.

In this paper, we fit the kinematics of the OB stars measured by \textit{Gaia} with the axisymmetric model and measure the $V_\mathrm{c}(R_{0})$ at the solar radius, $R_{0}$, and the slope of the circular velocity, $dV_\mathrm{c}(R_0)/dR$. We use the selection of the OB stars in the new \textit{Gaia} DR3 suggested by \citet{Drimmel2022gaia}. By applying the axisymmetric model to fit the kinematics of the OB stars, we assess how the circular velocity, $V_\mathrm{c}(R_{0})$, and its radial gradient, $dV_\mathrm{c}(R_0)/dR$, change depending on the size of the region to sample the tracer stars. The data and the model are summarised in Section \ref{sec:data} and \ref{sec:Model}, respectively. In Section \ref{sec:Results}, we report the results. Interestingly, we find that the gradient of the circular velocity changes depending on the size of the region to sample the tracers. In Section \ref{sec:discussion}, we compare our results with the literature and also the $V_\mathrm{c}(R_{0})$ measurement with Cepheids variables. We then compare our results with N-body simulation and discuss the impact of the Local arm on our measurements of $V_\mathrm{c}(R_{0})$. Section \ref{sec:summary} provides the summary of this study.

%%%%%%%%%%%%%%%%
\section{Data}
%%%%%%%%%%%%%%%%
\label{sec:data}

In this paper, we first select the young OB stars from the third data release \citep[Gaia $\textrm{DR3}$,][]{GaiaCollab2022b} of the European Space Agency's \textit{Gaia} mission \citep{GaiaCollab2016}. We select the OB stars following \citet{GaiaCollab2022c}, using the query shown in Appendix A of \citet{GaiaCollab2022c}. This query uses the information from the General Stellar Parameterizer from Photometry \citep[GSP-Phot,][]{Andrae2022gaia}. GSP-Phot provides stellar parameters based on \textit{Gaia}'s astrometry and low-resolution BP/RP photospectra, and the query also uses other information from Extended Stellar Parameterizer for Hot Stars \citep[ESP-HS,][]{Creevey2022gaia}. ESP-HS estimates stellar parameters and spectral types, especially for hot stars based on the BP/RP data or the combination of BP/RP spectra and high-resolution RVS spectra in addition to the parallax data. \citet{Drimmel2022gaia} used the following criteria to select the OB stars from this information. For the stars whose $\textrm{T}_{\textrm{eff}}$ is available only from GSP-Phot available, they select the stars with $\textrm{T}_{\textrm{eff}}$ from GSP-Phot of $\textrm{T}_{\textrm{eff}}> 10,000$ K as well as the spectral type label of ESP-HS of either “O”, “B” or “A”. For the stars whose $\textrm{T}_{\textrm{eff}}$ is available only from ESP-HS, $10,000<\textrm{T}_{\textrm{eff}}<50,000$ K are selected, because the stars with $\textrm{T}_{\textrm{eff}}>50,000$ K is likely an outlier due to poor model fits \citep{Fouesneau2022gaia}. For the stars whose $\textrm{T}_{\textrm{eff}}$ is available from both GSP-Phot and ESP-HS, they select $\textrm{T}_{\textrm{eff}}>8,000$ K for GSP-Phot and $\textrm{T}_{\textrm{eff}}>10,000$ K for ESP-HS, which helps to include the stars rejected by the criterion of $\textrm{T}_{\textrm{eff}}>50,000$ K from ESP-HS, but likely to be true hot stars. To remove the intrinsically faint stars, such as sub-dwarfs and white dwarfs, \citet{GaiaCollab2022c} also imposed the criterion of $(\pi/100)^5<10^{2-G+1.8(G_{\rm BP}-G_{\rm RP})}$, where $\pi$ is parallax and $G$, $G_{\rm BP}$ and $G_{\rm RP}$ are magnitudes at $G$, $G_{\rm BP}$ and $G_{\rm RP}$ band, respectively.

We further apply the astrometry quality cut, selecting the stars with Renormalized Unit Weight Error, $\texttt{RUWE}<1.4$ \citep{Lindegren2021b}, low parallax error ($\mathtt{parallax\_over\_error}>5$) and line-of-sight error and proper motion errors being less than 5~km~s$^{-1}$. 
The distances to the stars are obtained by simply the inverse of the \textit{Gaia}'s parallax measurements \citep{Lindegren2021a}. We also use the zero-point correction of parallax as suggested by \citet{Lindegren2021b}. Additionally, we applied the line-of-sight velocity correction for the hot stars recommended by \citet{Blomme2022}. Finally, we selected stars within the distance from the Sun of $\rm{D}<2$ kpc and the distance from the midplane of $|\rm{z}|<0.5$ kpc, assuming the Sun's height with respect to the midplane is $z_{\odot}=0.0208$ kpc \citep{Bennett2019}, to focus on the young stars in the disc.   

Following the work of \citet{Kawata2019} and \citet{Bovy2012}, we ignore the vertical motion or the thickness of the disc and consider the distance and the three component heliocentric velocities for our tracer stars projected onto the Galactic midplane (for more information on the model, see section \ref{sec:Model}). For heliocentric velocities, we use the line-of-sight velocities, $V_\mathrm{los}^\mathrm{helio}$, obtained from \texttt{radial\_velocity} of \textit{Gaia} $\textrm{DR3}$ including the above mentioned correction, and galactic longitudinal, $V_\textrm{glon}^\textrm{helio}$, and latitudinal, $V_\textrm{glat}^\textrm{helio}$, velocities, obtained from the proper motion of \texttt{pmra} and \texttt{pmdec} and \texttt{parallax} of \textit{Gaia} $\textrm{DR3}$. 

When we obtain the 2D velocities in parallel to the midplane from 3D velocities, we take into account the tilt of the b=0 plane from the plane of the Galactic disc, considering the height of the Sun from the midplane of $z_{\odot}=0.0208$ \unit{kpc} \citep{Bennett2019}, and the latitudinal proper motion of the Sun with respect to Sgr A*, $\mu_\mathrm{glat,\odot}$ \citep{ReidBrunthaler2020}. Note that we utilise 3D velocities obtained by \textit{Gaia}, compared to the previous studies of \citet{Bovy2012} who used $V_\mathrm{los}^\mathrm{helio}$ only and \citet{Kawata2019} who used $V_\mathrm{los}^\mathrm{helio}$ and $V_\mathrm{glon}^\mathrm{helio}$ only. 

%%%%%%%%%%%%%%%%%%%%%%%%%%%%%%%%%%%
\section{Model}
\label{sec:Model}
%%%%%%%%%%%%%%%%%%%%%%%%%%%%%%%%%%%
\subsection{Axisymmetric disc kinematic model}
\label{subsec:axisymmetricmodel}

\begin{figure*}
	\includegraphics[width= \textwidth]{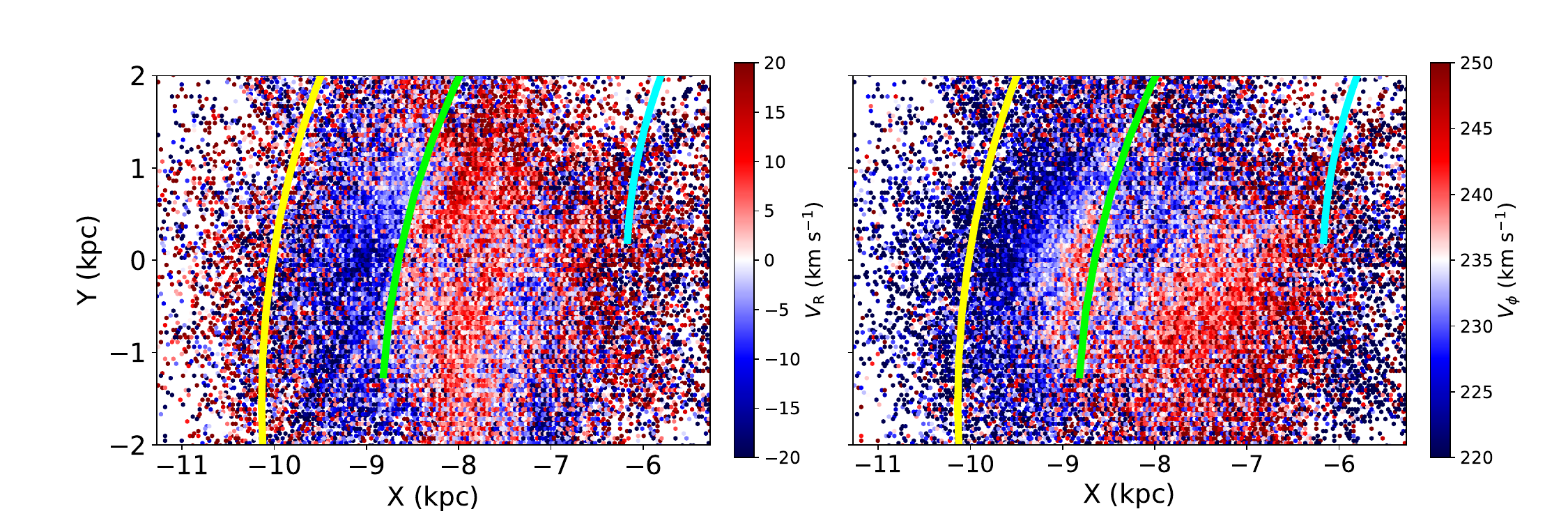}
    \caption{Galactocentric X-Y distribution of OB stars colour coded with $V_\mathrm{R}$ (left panel) and $V_{\phi}$ (right panel) overlaid with \citet{Reid2019} spiral arms. Perseus arm is in yellow, Local arm is in green, and Sagittarius arm is in cyan. We assume that the solar radial velocity, $V_\mathrm{R, \odot}= -9.65\pm0.43$ \unit{km.s^{-1}}, and the solar rotation velocity, $V_{\phi, \odot}= 247.74\pm2.22$ \unit{km.s^{-1}} to obtain our $V_\mathrm{R}$ and $V_{\phi}$ values from \textit{Gaia} DR3.}
    \label{XYdistribution}
\end{figure*}

We fit the observed mean and dispersion of $V_{\mathrm{los}}$ and $V_\mathrm{glon}$ in the Galactic rest frame with an axisymmetric Galactic disc model, and derive the posterior probability for the parameters of the models and solar motions by exploring likelihood of the model parameters with Markov Chain Monte Carlo (MCMC). The model is axisymmetric and assumes a Gaussian velocity distribution in the Galactic radial and rotation velocities with no correlation between these components, and zero mean radial velocity. These are simple model assumptions to describe the kinematics of the Galactic disc, but are sufficient to determine the local kinematic parameters of young stars in a relatively small region of the disc. 

In the axisymmetric disc model, we only need to consider the mean rotation velocity, $\overline{{V_{\phi}}}(R)$, and velocity dispersion in the Galactic rotation, $\sigma_{\phi}$, and radial, $\sigma_{R}$, directions because the mean radial velocity and the correlation between the two velocity components are assumed to be zero. The mean rotation velocity is calculated from asymmetric drift, $V_{a}$, as $\overline{V_{\phi}}(R) = V_\mathrm{c}(R)-V_\mathrm{a}(R)$, where $V_\mathrm{c}(R)$ is the circular velocity at radius $R$. We assume an exponential disc surface density profile, $\Sigma(R) \propto \exp \left(-R/h_\mathrm{R}\right)$, and an exponentially declining radial velocity, $\sigma_\mathrm{R} \propto \exp \left(-R/h_\mathrm{\sigma}\right)$, where $h_\mathrm{R}$ and $h_\mathrm{\sigma}$ are radial scale length for the surface density profile and velocity dispersion profile, respectively. Then the asymmetric drift can be calculated from the Jeans equation \citep[e.g.][]{BinneyTremaine2008} as,

\begin{equation}
\label{eq:axisymmetric}
    V_\mathrm{{a}}(R) = \frac{\sigma_\mathrm{R}^2(R)}{2 V_\mathrm{c}(R)} \left[ \left(\frac{\sigma_{\phi}}{\sigma_\mathrm{R}}\right)^{2} - 1 + R \left(\frac{1}{h_\mathrm{R}}+ \frac{2}{h_\mathrm{\sigma}}\right)\right].
\end{equation}

 \noindent Here, we assume $(\frac{\sigma_\mathrm{\phi}}{\sigma_\mathrm{R}})$ is constant for simplicity. We fix $h_\mathrm{R}=20$ \unit{kpc} and $h_\mathrm{\sigma}=20$ \unit{kpc} following \citet{Kawata2019}. However, we point out that the assumed scale length of the Milky Way disc is much larger than the conventional value, i.e. $h_\mathrm{R}=2.5-5$ \unit{kpc} \citep[e.g.][]{BlandHawthorn2016} because the scale length of radial velocity dispersion profiles for young stars are different from older ones. This assumption is also validated by \citet{Mackereth2017} who found that young stars have a  larger scale length of the surface density profile of $\sim6-7$ kpc. Since their sample of "young stars" are older than our OB stars, the younger OB stars can have a larger scale length. In any case, the asymmetric drift is less sensitive to the value of $h_\mathrm{R}$ or $h_\mathrm{\sigma}$ within these values, because for our very young stars, velocity dispersion is small and consequently the asymmetric drift is small \citep[see][for a further discussion]{Kawata2019}. We also assume that $V_\mathrm{c}$ follows a linear function of $R$ with the slope of $dV_\mathrm{c}(R_{0})/dR$ within the radial range of our tracer sample for simplicity. $dV_\mathrm{c}(R_{0})/dR$ is another parameter used to describe our axisymmetric disc model.

Furthermore, in the Galactic rest frame, the mean rotational velocity at the position of the star, $\overline{{V_{\phi}}}(R)$, can be projected onto the line-of-sight velocity, $V_\mathrm{los}$, as $V_\mathrm{m,los}= \overline{{V_{\phi}}}\sin(\phi+l)$, where $l$ is Galactic longitude and $\phi$ is the angle between the line from the Galactic center towards the Sun and the one towards the position of the star, positive in the direction of Galactic rotation. Additionally, we can derive the projected longitudinal velocity from the rotation velocity of the axisymmetric model as $V_\mathrm{m,glon}= \overline{{V_{\phi}}}\cos(\phi+l)$.

On the other hand, the observational data provide the line-of-sight velocity, $V^{\mathrm{helio}}_{\mathrm{los}}$, Galactic longitudinal velocity, $V^{\mathrm{helio}}_{\mathrm{glon}}$, with respect to the Sun's motion. Using the solar radial velocity, $V_\mathrm{R, \odot}$ (outward motion is positive), and rotational velocity, $V_{\phi, \odot}$ (clockwise rotation is positive), we convert  $V^\mathrm{helio}_\mathrm{los}$ and $V^\mathrm{helio}_\mathrm{glon}$ to the Galactocentric rest frame as follows:

\begin{equation}
    V_\mathrm{o, los}=  V^\mathrm{helio}_\mathrm{los}-V_\mathrm{R,\odot}\cos(l)-V_{\phi, \odot}\sin(l),
\end{equation}

\begin{equation}
    V_\mathrm{o, glon}=  V^\mathrm{helio}_\mathrm{glon}-V_\mathrm{R,\odot}\sin(l)+V_{\phi, \odot}\cos(l).
\end{equation}

\noindent These values can then be compared with the expected velocity distribution of $V_\mathrm{m,los}$ and $V_\mathrm{m,glon}$ from the model. Hence, fitting our simple axisymmetric disc model to the observed young sample tracer stars requires us to explore seven model parameters, $\theta_{m}= \{ V_\mathrm{c}(R_{0}), V_{\phi, \odot}, V_\mathrm{R,\odot},\sigma_{R}(R_{0}), X^{2}, R_{0}, dV_\mathrm{c}(R_{0})/dR \}$, where $X^{2}=(\sigma_\mathrm{\phi}/\sigma_\mathrm{R})^2$.

\subsection{MCMC parameter probabilities}
\label{subsec:MCMC}

Following \cite{Kawata2019}, we use Bayes’ theorem to find the marginalized probability distribution function of our model parameters as follows:

\begin{equation}
    p(\theta_\mathrm{m}|\mathcal{D})=  \mathcal{L}(\mathcal{D}|\theta_\mathrm{m}) \times \mathrm{Prior},
\end{equation}

\noindent where $\mathcal{D} = (V^{\mathrm{helio}}_{\mathrm{los}}, V^{\mathrm{helio}}_{\mathrm{glon}})$ represents the whole observed data set values in the star sample, and $\theta_\mathrm{m}$ represents the combination of all the model parameters described in Section \ref{subsec:axisymmetricmodel}. We run the MCMC using this $p(\theta_\mathrm{m}|\mathcal{D})$. The likelihood function is 

\begin{equation}
    \mathcal{L}(\mathcal{D}|\theta_\mathrm{m})= \prod_\mathrm{i}^{N} \frac{1}{2\pi|\mathbf{C}_\mathrm{i}|^{1/2}} \mathrm{exp}(-0.5\mathbf{X}_\mathrm{i}^\mathrm{T} \mathbf{C}_\mathrm{i}^{-1} \mathbf{X}_\mathrm{i} ),
\end{equation}

where 

\begin{equation*}
\mathbf{X}_\mathrm{i} = 
\begin{pmatrix}
V_\mathrm{o, los,i} - V_\mathrm{m, los,i}  \\
V_\mathrm{o, glon,i} - V_\mathrm{m, glon,i} 
\end{pmatrix}
.
\end{equation*}

\noindent Here, $V_{ \mathrm{o}, \mathrm{los,i}}$ and $V_{\mathrm{o}, \mathrm{glon,i}}$ are the observed line-of-sight and longitudinal velocity in the Galactic rest frame for star i, respectively, and $V_{ \mathrm{m}, \mathrm{los,i}}$ and $V_{ \mathrm{m}, \mathrm{glon,i}}$ are the expected line-of-sight and longitudinal velocity in the Galactic rest frame from the axisymmetric model at the location of star i, respectively. The covariance matrix, $\mathbf{C}_\mathrm{i}$, is $\mathbf{C}_\mathrm{i}$ = $\mathbf{A}_\mathrm{i}\mathbf{S}_\mathrm{i}\mathbf{A}_\mathrm{i}^{T}$, where

\begin{equation*}
\mathbf{A}_\mathrm{i} = 
\begin{pmatrix}
 -\cos(\phi_\mathrm{i} + l_\mathrm{i}) & \sin(\phi_\mathrm{i} + l_\mathrm{i})   \\
  \sin(\phi_\mathrm{i} + l_\mathrm{i}) & \cos(\phi_\mathrm{i} + l_\mathrm{i})
\end{pmatrix},
\end{equation*}

and

\begin{equation*}
\mathbf{S}_\mathrm{i} = 
\begin{pmatrix}
 \sigma_\mathrm{R_{i}}^{2} & 0   \\
 0 & \sigma_\mathrm{\phi_{i}}^{2}
\end{pmatrix}.
\end{equation*}

To take into account the uncertainties of the observational data, we generate 1,000 Monte Carlo (MC) samples of parallax, proper motions in RA and DEC and line-of-sight velocity for each stars from Gaussian probability distribution with their observed values and errors. When we sample these values, we also take into account the correlation among parallax and RA and DEC proper motions using \texttt{parallax\_pmra\_corr}, \texttt{parallax\_pmdec\_corr} and \texttt{pmra\_pmdec\_corr} in \textit{Gaia} DR3.
Then, these samples are converted to the observed heliocentric velocities. We compute likelihood for these MC sampled data points for each star and take the average of them. This can be described as

\begin{equation}
\label{eqn:likelihood}
    \mathcal{L}(\mathcal{D}|\theta_\mathrm{m})= \prod_{i}^{N} \int p(\mathcal{D}_{\mathrm{obs,i}}|\mathcal{D}_{\mathrm{true,i}}) \mathcal{L}(\mathcal{D}_{\mathrm{true,i}}|\theta_{m}) d\mathcal{D}_{\mathrm{true,i}},
\end{equation}

\noindent where $\mathcal{D}_{\mathrm{obs,i}}$ are the observed values for the \textit{i}-th star and $\mathcal{D}_{\mathrm{true,i}}$ are the error-free, true values of these observables predicted by the model. We approximate the integral of equation (\ref{eqn:likelihood}) with the MC sampling as described above.

We note that $R_{0}$ is not well constrained by our observable data and we introduce a Gaussian prior for $R_{0}$ as follows:

\begin{equation}
\label{eqn:GaussianPrior}
    \mathrm{Prior}(R_{0}) = \frac{1}{\sqrt{2\pi\sigma^{2}_\mathrm{{R_{0, \mathrm{prior}}}}}} \exp{\left(-\frac{(R_{0}-R_{0,\mathrm{prior}})^2}{2\sigma_\mathrm{R_{0}, \mathrm{prior}}^2}\right)},
\end{equation}

\noindent where $R_{0, \mathrm{prior}}= 8.275$ \unit{kpc}, and $\sigma_\mathrm{{R_{0}}, \mathrm{prior}}= 0.034$ \unit{kpc}, the strong priors obtained from \citet{GravityCollab2021}. Furthermore, we use a Gaussian prior for the angular velocity of the Sun, $\Omega_{\odot }=V_{\phi,\odot}/R_{0}$, with $\Omega_{\odot} = 30.32 \pm 0.27$ \unit{km.s^{-1}.kpc^{-1}}, because it is well constrained by \cite{ReidBrunthaler2020}.
We use the \texttt{emcee} package \citep{Goodman2010, Foreman2013emcee} MCMC sampler to our data. Our MCMC model is defined using the following specifications: 200 walkers, 1000 chains for each walker, and 200 chains for burn-in.

%%%%%%%%%%%%%%%%%
\section{Results}
\label{sec:Results}
%%%%%%%%%%%%%%%%%

Fig. \ref{XYdistribution} shows the X-Y distribution of the OB stars in the Galactocentric frame obtained from the selection criteria described in Section \ref{sec:data}, colour-coded with $V_\mathrm{R}$ and $V_\mathrm{\phi}$, and
overlaid with the position of the spiral arms from \citet{Reid2019}. Here, the X and Y axes are the Galactocentric coordinate with the Galactic center to be $(\textrm{X, Y})=(0, 0)$ and the location of the Sun is $(\mathrm{X, Y}) = (-R_\mathrm{0}, 0)$. Y-axis is parallel to the direction of the Sun's rotation velocity, i.e. $V_\mathrm{Y} = V_\mathrm{\phi,\sun}$. To plot the distribution of OB stars in this coordinate, we assume $R_\mathrm{0}=8.28~\mathrm{kpc}$. We also assume that the solar radial velocity, $V_\mathrm{R, \odot}= -9.65$ \unit{km.s^{-1}}, and the solar rotation velocity, $V_{\phi, \odot}= 247.74$ \unit{km.s^{-1}} used to calculate our $V_\mathrm{R}$ and $V_{\phi}$ values from \textit{Gaia} DR3. As explained below, these values are obtained from our best fit model for the OB stars data within $\mathrm{D}<2$ \unit{kpc}.

\begin{figure}
	\includegraphics[width=\columnwidth]{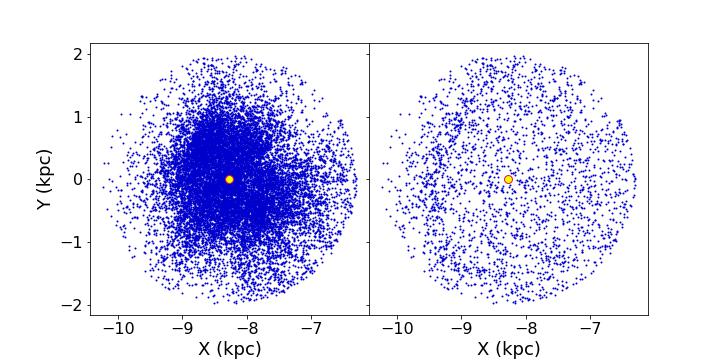}
    \caption{X-Y distribution of OB stars within $\mathrm{D}<2.0~\mathrm{kpc}$ (left panel) and after random sampling of OB stars in the (right panel). The yellow dot represents the position of the Sun. } 
    \label{randomsampling}
\end{figure}

Fig. \ref{XYdistribution} displays that our sample of the OB stars are concentrated within $\mathrm{D}<2~\mathrm{kpc}$ from the Sun, while the number of the sample is significantly lower than outside of $\mathrm{D}\sim2$ kpc. The velocity structure is obviously not axisymmetric, which is a challenge to our axisymmetric model. These structures also coincide with the location of the spiral arm. However, we first apply the simplest model, and later we will discuss the impact of the non-axisymmetric structure in Section \ref{sec:discussion}. Because the number density of stars drops outside of $\mathrm{D}=2~\mathrm{kpc}$, we use data within $\mathrm{D}<2~\mathrm{kpc}$ for our fitting of the axisymmetric disk model.  The left panel of Fig. \ref{randomsampling} illustrates the distribution set of OB stars within  $\mathrm{D}<2~\mathrm{kpc}$ from the Sun. The OB stars are not evenly distributed and are mainly concentrated around $\mathrm{D}<1~\mathrm{kpc}$ from the Sun. In order to eliminate any sampling biases, and to sample the likelihood homogeneously in the region of interest, we utilise the homogenisation process inspired by \citet{Kawata2019}. In this process, we randomly choose 2,500 OB stars that are located within $2~\mathrm{kpc}$ from the Sun's position. This is done by a repeating process of selecting the nearest star to a randomly chosen location, but discarding any stars that were previously selected before proceeding to pick the next closest star based on the randomly chosen location. This is depicted in the right panel shown in Fig. \ref{randomsampling}, where the stars look more evenly distributed. We set a limit of 2,500 OB stars because this is a good compromised choice to maximise the number of the sample of stars, but selecting them in a somewhat uniform and random manner.

\begin{figure*}
	\includegraphics[width= \textwidth]{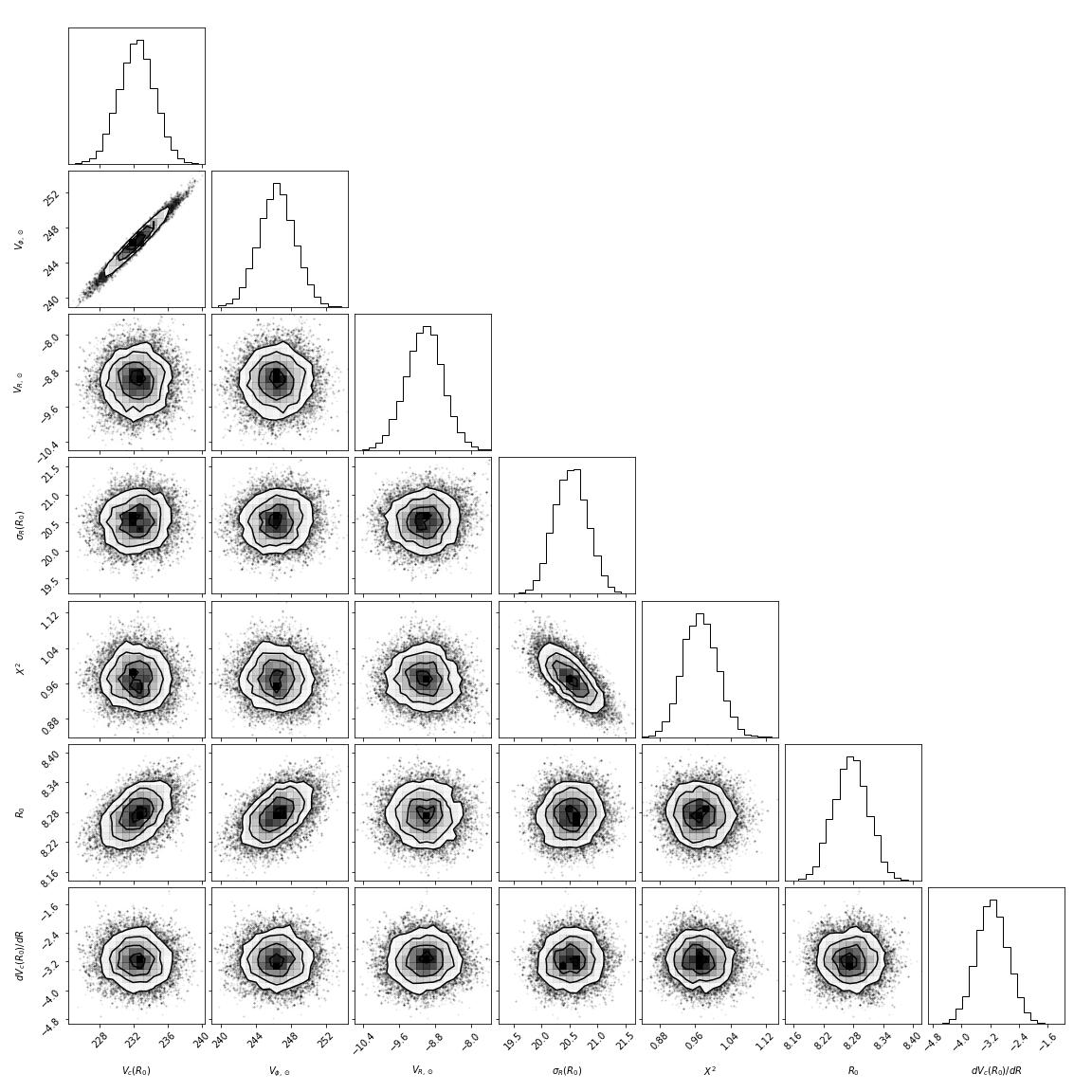}
    \caption{ Marginalized posterior probability distribution of our model parameters for $\mathrm{D}<2$ kpc sample fitting.}
    \label{modelparam(OBStars)}
\end{figure*}

We apply our axisymmetric disk model fit to the OB stars data obtained in the right panel of Fig. \ref{randomsampling}. The marginalized posterior probability distribution of our parameters are shown in the corner plot of Fig. \ref{modelparam(OBStars)}. We consider the mean and dispersion of the posterior probability distribution as the best fit parameter value and the uncertainty, and they are summarized in Table \ref{tab:dataresults}.

\begin{table*}
	\centering
	\caption{MCMC Fitting results (mean and dispersion of all parameters).}
	\label{tab:dataresults}
	\begin{tabular}{ lcccccccr } 
		\hline
		 & $\mathrm{D}<0.5$ \unit{kpc} & $\mathrm{D}<2.0$ \unit{kpc} &  Cepheids    \\
		\hline
		$V_\mathrm{c}(R_0)$ \unit{(km.s^{-1})} & $238.30\pm2.43$ & $233.95\pm2.24$ & $239.16\pm2.52$   \\
		$V_{\phi,\odot}$ \unit{(km.s^{-1})} & $250.61\pm2.51$ & $247.74\pm2.22$ & $250.55\pm2.44$ \\
		$V_\mathrm{R,\odot}$ \unit{(km.s^{-1})} & $-11.34\pm0.35$ & $-9.65\pm0.43$ & $ -8.77\pm1.20$  \\
		$\sigma_\mathrm{R}(R_0)$ \unit{(km.s^{-1})} & $14.83\pm0.27$ & $20.81\pm0.28$  & $14.34\pm0.85$  \\
		$(\sigma_\phi/\sigma_\mathrm{R})^{2}$  & $0.59\pm0.03$ & $1.05\pm0.04$  & $0.38\pm0.07$  \\
		$R_0$ \unit{(kpc)} & $8.28\pm0.03$ & $8.28\pm0.04$ & $8.28\pm0.03$ & \\
		$dV_\mathrm{c}(R_0)/dR$ \unit{(km.s^{-1}.kpc^{-1})} & $1.82\pm1.17$ & $-3.31\pm0.48$ & $-3.47\pm0.79$   \\
            N  & 1600 & 2500 & 154   \\
		\hline
	\end{tabular}
\end{table*}

Fig. \ref{modelparam(OBStars)} shows that the total rotation velocity of the Sun, $V_\mathrm{\phi, \odot}$, is strongly correlated with $V_\mathrm{c}(R_{0})$. This is because the Sun rotates similar speed to the circular velocity. If we subtract the circular velocity, $V_\mathrm{\phi, \odot}-V_\mathrm{c}(R_{0})$, we do not see any correlation.

%(Please confirm. Please make a plot of PDF of $V_\textrm{\phi, \sun}-V_\textrm{c}(R_0)$ vs Vc.)

\begin{figure}
	\includegraphics[width= \columnwidth]{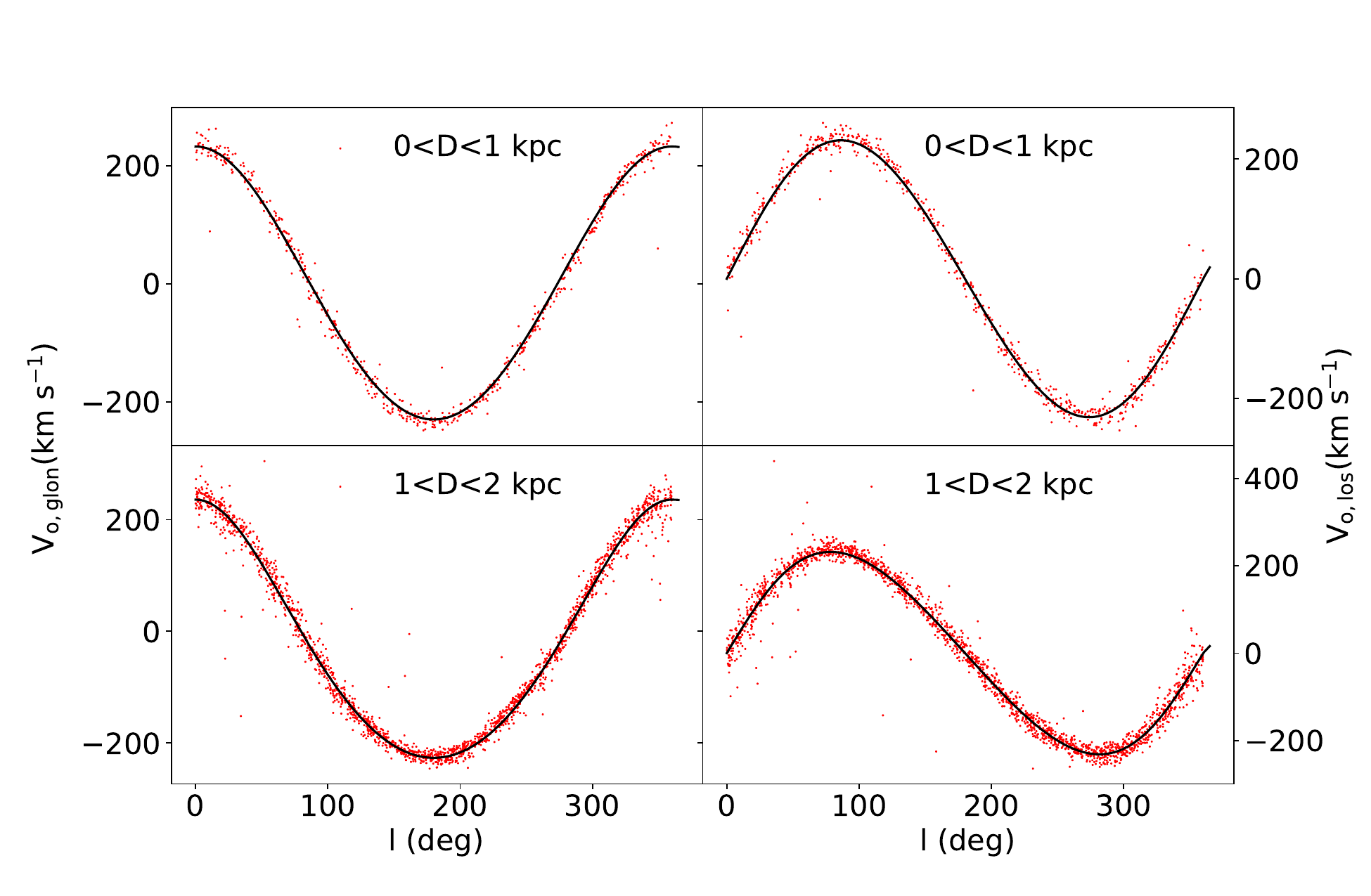}
    \caption{Observed Galactic longitudinal velocity, $V_\mathrm{o, glon}$, (left) and observed line-of-sight velocity, $V_\mathrm{o, los}$, (right) as a function of Galactic longitude, $l$, of our OB star sample with distances between 0 and 1 kpc (upper), and between 1 and 2 kpc (lower). The solid black line in each panel represents the result from our best-fitting model using $\mathrm{D}=0.5$ \unit{kpc}, and $\mathrm{D}=1.5$ \unit{kpc} for the upper and lower panel, respectively.}
    \label{lvsVglonVlos}
\end{figure}

%Please explain what Fig. 4 are showing and demonstrate that your model describes the observational data well. 
Fig. \ref{lvsVglonVlos} presents the trends of $V_\mathrm{o, glon}$ and $V_\mathrm{o, los}$ as a function of Galactic longitude, $l$, based on the kinematics of selected OB stars for our MCMC fitting. The figure also includes the trends obtained from the best-fitting model at $\mathrm{D}=0.5$ \unit{kpc} in the upper panel, and at $\mathrm{D}=1.5$ \unit{kpc} in the lower panel. The observed kinematics of OB stars is overall consistent with our best-fit model.

Table \ref{tab:dataresults} shows that we obtain $V_\mathrm{c}(R_{0}) = 233.95\pm2.24$ \unit{km.s^{-1}}. This is similar to what are obtained in literature, and in Section \ref{subsec:comparisonwithliterature}, we discuss the comparison of our results with the literature values. We also find a negative $V_\mathrm{c}(R_{0})$ gradient of $dV_\mathrm{c}(R_{0})/dR= -3.31\pm0.48$ \unit{km.s^{-1}.kpc^{-1}}. We also obtain $R_{0}=8.28\pm0.04$ \unit{kpc}. However, this is strongly constrained by our prior of $R_{0}= 8.275\pm0.034$ \unit{kpc} \citep{GravityCollab2021}. 

Although our axisymmetric model fit works well, Fig. \ref{XYdistribution} clearly shows the non-axisymmetric velocity structure. It would be important to assess how the sample selection affects the results \citep[e.g.][]{Nitschai2021}. To this end, we also apply our axisymmetric model fit to the data within $\mathrm{D}<0.5$~kpc. In this region, we also implement the same homogenisation technique shown in Fig. \ref{randomsampling} to the data set and randomly select 1,600 stars. The summary of the best-fit parameter values and uncertainty after our MCMC sampling is shown in Table\ref{tab:dataresults}. We obtain slightly higher value of $V_\mathrm{c}(R_{0}) = 238.30\pm2.43$ \unit{km.s^{-1}}, which is about $2\sigma$ higher than $V_\mathrm{c}(R_{0})$ found from the $D<2$ \unit{kpc} sample fit. Interestingly, we find a positive $dV_\mathrm{c}(R_{0})/dR= 1.82\pm1.17$ \unit{km.s^{-1}.kpc^{-1}} compared to a negative slope obtained in the $\mathrm{D}<2$ \unit{kpc} sample fit, although the uncertainty is relatively high. This indicates that the velocity structures change even in a kpc scale, which is also seen in Fig. \ref{XYdistribution} and likely induced by the spiral arm \citep[see also][]{Bobylev2021}. In this case, as shown in Fig. \ref{XYdistribution}, the velocity structure within $\mathrm{D}<2$ kpc is likely caused by the Local arm.

\begin{figure}
        \includegraphics[width=\columnwidth]{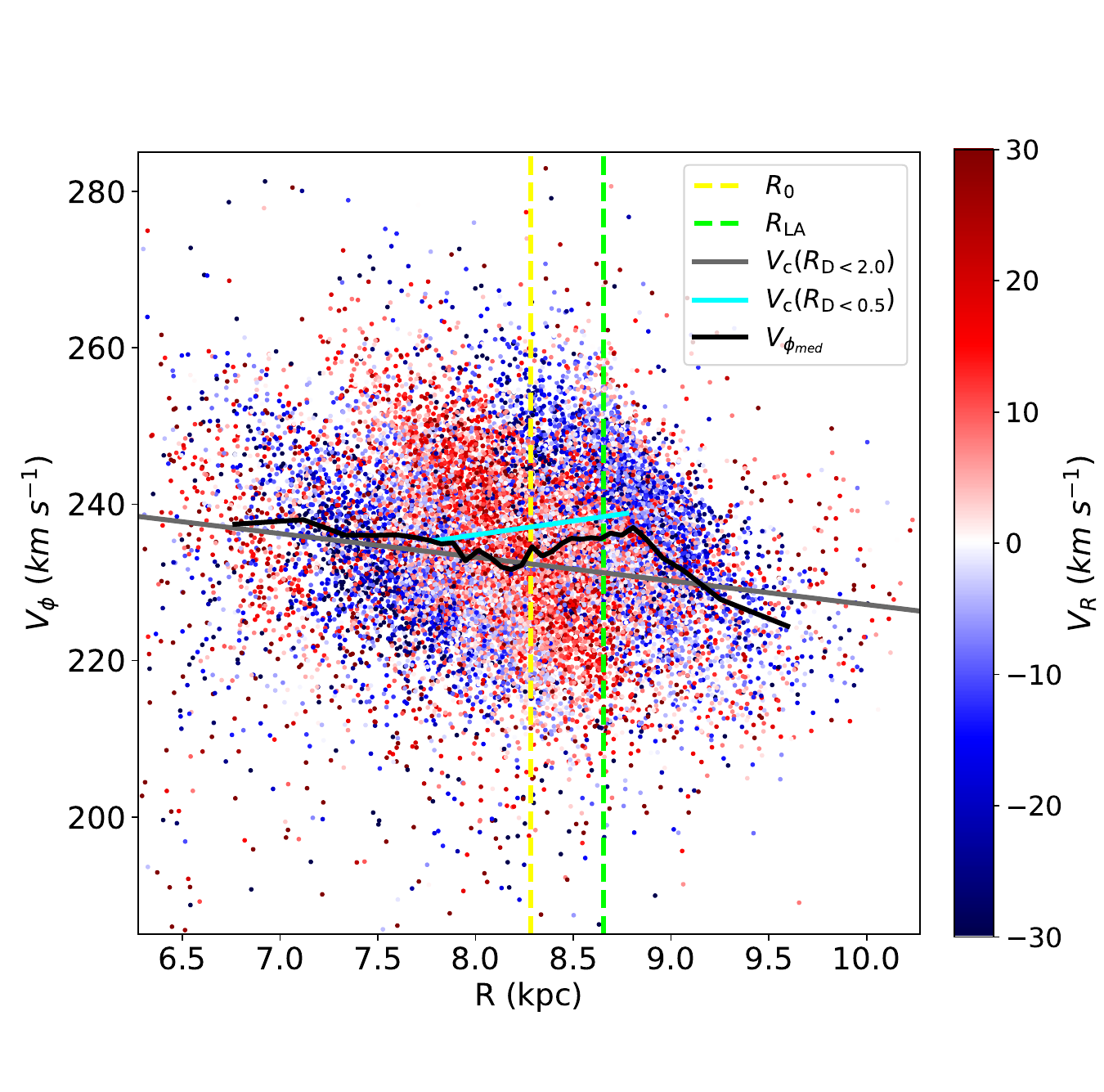}
    \caption{Map of $V_\mathrm{\phi}$ distribution of OB stars at different $R$ positions colour-coded with $V_\mathrm{R}$. The grey line represents the best fit $V_\mathrm{c}(R)$ result from the $\mathrm{D}<2.0$ \unit{kpc} OB star sample. The cyan line represents the best fit $V_\mathrm{c}(R)$ result from the $\mathrm{D}<0.5$ \unit{kpc} OB star sample. The black line represents the binned median of $V_{\phi}$ in the entire OB population. The yellow vertical dashed line represents $R_{0}=8.28$ \unit{kpc}. The green vertical dashed line represents the location of the Local arm at  $R_\mathrm{LA}=8.65$ \unit{kpc}.} 
    \label{r-vphi(OBstars)}
\end{figure}

To see how the rotation velocity of the OB stars are influenced by the location of the Local arm, Fig. \ref{r-vphi(OBstars)} shows $R-V_\mathrm{{\phi}}$ distribution of all the OB stars. Here, the dots are colour-coded with $V_\mathrm{R}$. The black line indicates the median $V_\mathrm{\phi}$ of the OB stars as a function of $R$. Here, $R_{0}=8.28$ \unit{kpc} is set to be the best fit parameter obtained from the sample of $\mathrm{D}<2$ \unit{kpc}. We find an interesting relation between $V_\mathrm{\phi}$ and $V_\mathrm{R}$ depending on $R$. Overall $V_\mathrm{\phi}$ decreases with the radius, which is consistent with the negative $dV_\mathrm{c}(R_{0})/dR$ slope obtained from the OB stars sample within $\mathrm{D}<2$ kpc, shown in the grey line. However, in $[-0.5+R_{0}<R<R_{0}+0.5]$ kpc, the median $V_\mathrm{\phi}$ increases with radius, which agrees with the positive $dV_\mathrm{c}(R_{0})/dR$ found from the $\mathrm{D}<0.5$ \unit{kpc} sample shown in the cyan line. In the same radial range, $V_\mathrm{R}$ changes from positive value (outward motion) to negative value (inward motion) as the radius increases. This is a typical velocity trend seen around the spiral arm in the numerical simulations \citep[e.g.][]{Grand2012b, Baba2013} whose spiral arms are co-rotating, dynamic and transient. \citet{Kawata2014} show this trend using a Milky Way-sized spiral galaxy N-body and smooth particle hydrodynamics (SPH) simulation where inside of the spiral arm both stars and gas are rotating slower and moving outward, while the gas and stars outside of the arm are rotating faster and moving inward. This motion is also known to drive the radial migration of stars and gas because the spiral arms are co-rotating \citep{Grand2012b, Grand2016}. Interestingly, this motion is detected in the ionized gas motion of an external galaxy, NGC 6754 \citep{SanchezMenguiano2016}. 

This means that the Local arm significantly influences the motion of very young stars, such as OB stars. The Local arm is located at a slightly larger radius than $R_{0}$ where $V_\mathrm{\phi}$ starts decreasing with the radius. The location of the Local arm at Y=0 in Fig. \ref{XYdistribution} is shown with the vertical red dashed line in Fig. \ref{r-vphi(OBstars)}. This is visually consistent with the rotation velocity distribution of the gas in the lower panel of Fig. 4 of \citet{Kawata2014}. They are compared with the motion of the high-mass star-forming region around the Perseus arm. However, this motion around the Local arm infers that the Local arm, which is often considered to be a minor spur, could be a strong spiral arm, and clearly influences the motion of the stars around it. We will discuss this more in Section \ref{sec:discussion}, and discuss the impact on the estimates of $V_\mathrm{c}(R_{0})$.

%%%%%%%%%%%%%%%%%%%%%%%%%
\section{Discussion}
\label{sec:discussion}
%%%%%%%%%%%%%%%%%%%%%%%%%
\subsection{Comparison with literature}
\label{subsec:comparisonwithliterature}

\begin{figure*}
	\includegraphics[width= 0.6 \textwidth]{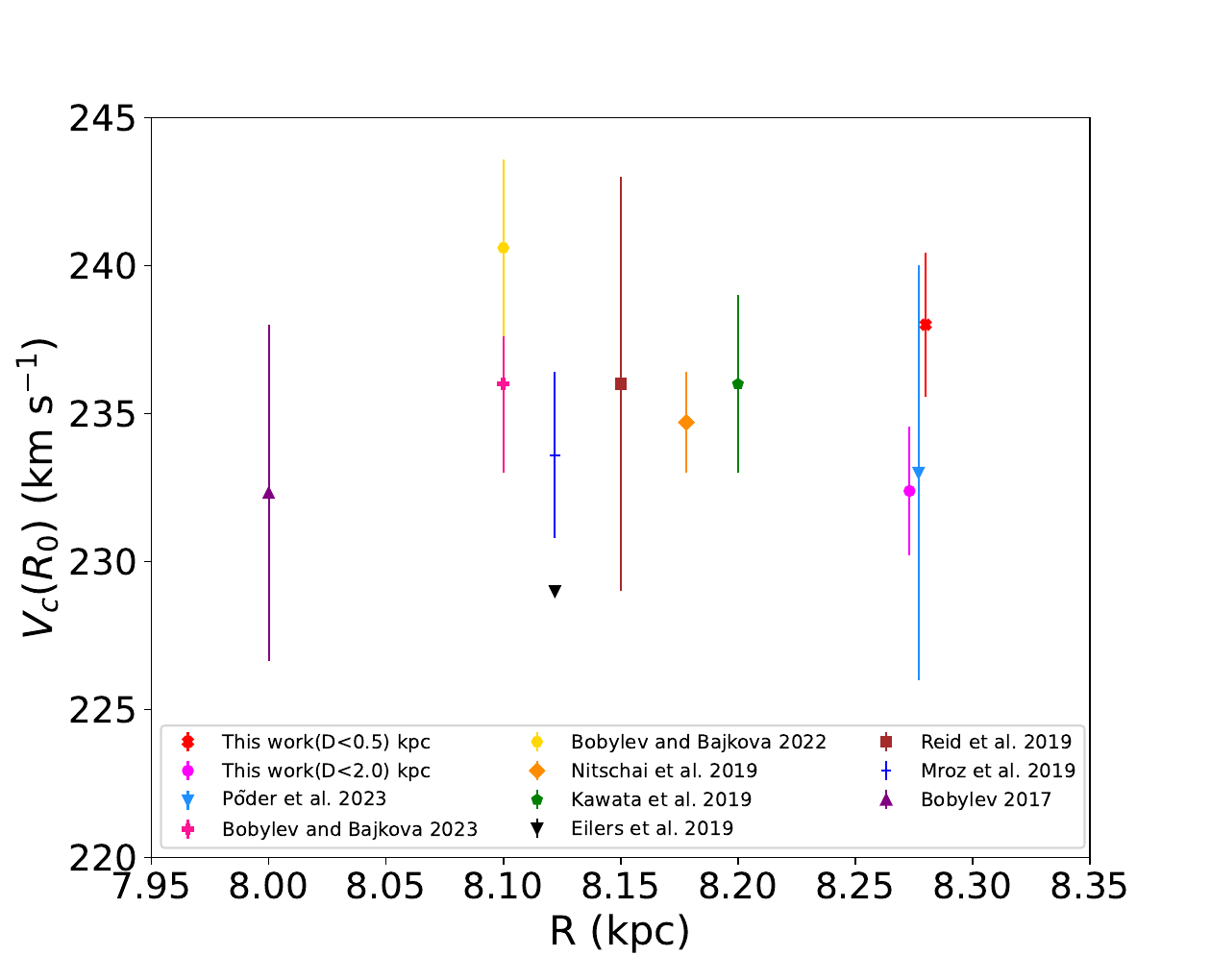}
    \caption{Measurements of the circular velocity values of the Milky Way at $R_{0}$ from various literature.}
    \label{resultsVsliterature}
\end{figure*}

We obtained different $V_\mathrm{c}(R_{0})$ by fitting the OB stars within $\mathrm{D}<0.5$ \unit{kpc} and $\mathrm{D}<2$ \unit{kpc}. In this section, we compare these $V_\mathrm{c}(R_{0})$ with the circular velocity at the solar radius obtained by the previous works. The summary of the comparison is shown in Fig. \ref{resultsVsliterature}. 

\citet{Mroz2019} examined Cepheid sourced from \citet{Skowron2019b}. They obtained $V_\mathrm{c}(R_{0})= 233.6\pm2.8$ \unit{km.s^{-1}} and $R_{0}= 8.27\pm0.1$ kpc from their Cepheid sample. This value was derived using a prior value of $R_0=8.112\pm0.031$ kpc obtained from the work of \citet{Abuter2018gravity}. It is important to note that the prior value used in \citet{Mroz2019} is slightly different from the one we are using, which is an updated $R_{0}$ measurement of \citet{GravityCollab2021}. 

\citet{Bobylev2017} obtained $V_\mathrm{c}(R_{0})= 231\pm6$ \unit{km.s^{-1}} using 249 Cepheids assuming $R_{0}= 8$ kpc to be their prior. This value was later updated in \citet{Bobylev2021} to be $V_\mathrm{c}(R_{0})= 240\pm3$ \unit{km.s^{-1}} by utilising 788 Cepheid samples from \citet{Skowron2019b} for the calculated value of $R_{0}= 8.27\pm0.1$ kpc, but not plotted in Fig. \ref{resultsVsliterature}. Because this is superseded by their more recent work of \citet{BobylevBajkova2023}, who found $V_\mathrm{c}(R_{0})= 236\pm3$ \unit{km.s^{-1}} using Cepheid as the tracer population. 

It is important to note that the studies by \cite{Bobylev2017}, \citet{Mroz2019}, \cite{Bobylev2021}, and \cite{BobylevBajkova2023} assume that the mean rotation velocity of young stars is close to the circular velocity and that asymmetric drift is zero. In contrast, our study takes into account the effect of asymmetric drift, even though it is small. The mean value of the asymmetric drift for our OB stars is $V_{\mathrm{a}}=1.22$ \unit{km.s^{-1}}.

\citet{BobylevBajkova2023} also found $V_\mathrm{c}(R_{0})= 240.6\pm3$ \unit{km.s^{-1}} for 9,750 OB stars. The proper motion and trigonometric parallaxes of the OB stars were taken from \citet{Xu2018} using \textit{Gaia}'s Early Data Release 3 (EDR3) catalogue. The value of $R_{0}$ was assumed to be $8.1\pm0.1$ kpc, which was taken from a review by \citet{BobylevBajkova2021}.

\citet{Eilers2019} derived the value of $V_\mathrm{c}(R_{0})= 229\pm0.2$ \unit{km.s^{-1}} by using red giant stars obtained from The Apache Point Observatory Galactic Evolution Experiment (APOGEE) Data Release 14 (DR14), as well as the complete photometric information in the $G$, $G_{\mathrm{BP}}$, and $G_{\mathrm{RP}}$ band from \textit{Gaia} Data Release 2 (DR2) \citep[DR2,][]{GaiaCollab2016}. They employed the axisymmetric model and assumed ${h}_\mathrm{R}=3$ kpc from \citet{BlandHawthorn2016}. $h_\mathrm{\sigma}$ was also derived to be 21 kpc for their tracer population. They used the same equation as our equation (\ref{eq:axisymmetric}). \citet{Poder2023} have also followed the same method as \citet{Eilers2019} and have found $V_\mathrm{c}(R_{0})= 233.0\pm6.0$ \unit{km.s^{-1}} for $R_{0}= 8.277$ kpc by using red giants as well.  

\citet{Nitschai2021} found $V_\mathrm{c}(R_{0})= 234.7\pm1.7$ \unit{km.s^{-1}} by combining two data sets, one was from giant stars from \textit{Gaia} Early Data Release 3 \citep[EDR3,][]{GaiaCollab2021}, and one was the red giant branch stars from APOGEE and \textit{Gaia} obtained from \citet{Hogg2019}. They assumed $R_{0}= 8.178$ kpc obtained from \citet{Abuter2018gravity}, and vertical placement of the Sun from the midplane of $z_{\odot}=0.02$ kpc from \citet{Joshi2007}.
 
\citet{Reid2019} found $V_\mathrm{c}(R_{0})= 236\pm7$ \unit{km.s^{-1}} and $R_{0}= 8.15\pm0.15$ kpc using a sample of massive star forming regions. The measurements of their data come from the Bar and Spiral Structure Legacy (BeSSeL) Survey and the Japanese Very Long Baseline Interferometry (VLBI) Exploration of Radio Astrometry (VERA) project. \citet{Reid2019} adopted a loose prior on the $V_{\odot}$ of the solar motion and for the average peculiar motion of the massive star-forming region stars.  

By studying a sample of Cepheids, \citet{Kawata2019} applied an axisymmetric model to the population of these stars and derived a value of $V_\mathrm{c}(R_{0})= 236\pm3.0$ \unit{km.s^{-1}}. The study utilized a prior assumption of $R_0= 8.2\pm0.1$ kpc and $\Omega_{\odot} = 30.24 \pm 0.12$ \unit{km.s^{-1}.kpc^{-1}} from \citet{ReidBrunthaler2004}. Since we follow a similar methodology as \citet{Kawata2019}, we can compare all our obtained parameter values with their results. The value of $V_\mathrm{c}(R_{0})$ falls within the uncertainty range of what is obtained in our model. In \citet{Kawata2019}, the value of $V_\mathrm{R,\odot}=-7.7\pm0.9$ \unit{km.s^{-1}} is much slower than our value of $V_\mathrm{R,\odot}=-9.65\pm0.43$ \unit{km.s^{-1}} reported in Table \ref{tab:dataresults}. However, \cite{Schonrich2012} reports $V_\mathrm{R,\odot}=-14$ \unit{km.s^{-1}} of dwarf stars within $\mathrm{D}<4$ \unit{kpc}.  The data was collected using the Sloan Digital Sky Survey (SDSS) DR8 \citep{Eisenstein2011} and the stellar spectra from SEGUE \citep{Yanny2009}.  It is important to note that the value of $V_\mathrm{R,\odot}$ is heavily influenced by the mean radial velocities of the tracer population, $\overline{V_\mathrm{R, \mathrm{stars}}}$, and the true value of the radial velocity of the Sun, $V_\mathrm{R,\odot, \mathrm{ true}}$, as $V_\mathrm{R,\odot}= V_\mathrm{R,\odot, \mathrm{true}}-\overline{V_\mathrm{R,\mathrm{stars}}}$. Hence, the differences in $V_\mathrm{R,\odot}$  may indicate the different systematic mean radial motion for the different tracer samples, perhaps due to the spiral arm. 

\citet{Kawata2019} reports $V_{\phi,\odot}=252.4\pm10.7$ \unit{km.s^{-1}}, which falls within the uncertainty range of our value. Additionally, \citet{Kawata2019} obtained a lower velocity dispersion, $\sigma_\mathrm{R}(R_0)=13.0\pm0.6$ \unit{km.s^{-1}} for Cepheid, than what is reported for OB stars. Nevertheless, the observed velocity dispersion corresponds more closely to the velocity dispersion we identified in OB stars within $\mathrm{D}<0.5$ kpc. Furthermore, the value of $(\sigma_\phi/\sigma_\mathrm{R})^{2}=0.61\pm0.08$ in \citet{Kawata2019} aligns more with the results obtained from our sample in $\mathrm{D}<0.5$ kpc. This could suggest that our sample at $\mathrm{D}<2$ kpc is substantially contaminated by older and higher velocity dispersion stars or the error reported in \textit{Gaia} DR3 for the OB stars might be understated. On the other hand, \citet{Babusiaux2023} suggested that the line-of-sight velocity of \textit{Gaia} DR3 is underestimated especially for hotter stars. Hence, we have increased the uncertainty of the line-of-sight velocity by a factor of two to test whether the higher uncertainty would affect our model results. However, we do not find any impact on our results. Hence, it is unlikely that the underestimated uncertainty of the line-of-sight velocity is the cause of this difference. Nevertheless, the generally consistent findings from the sample at $\mathrm{D}<2$ kpc suggest that the average trend of the results holds relevance. \citet{Kawata2019} also obtained $dV_\mathrm{c}(R_0)/dR=-3.6\pm0.5$ \unit{km.s^{-1}.kpc^{-1}} which is consistent with the OB star sample within $\mathrm{D}<2$ kpc.

\subsection{Comparison with Cepheids}
%Comparison with Cepheids data 

To benchmark our OB stars results, we run the same MCMC model on Cepheids. The Cepheid sample used in our study is sourced from \cite{Skowron2019a}, which originates from the Optical Gravitational Lensing Experiment (OGLE) project \citep{Udalski2015ogle}, a comprehensive survey focused on discovering and classifying variable stars in the Galactic disc and Galactic center. The distances of individual Cepheids in this data set were determined using mid-infrared photometry obtained from the Spitzer and Wide-field Infrared Survey Explorer (WISE) satellites. These distances were corrected for interstellar dust using methods described in \cite{Skowron2019b}, along with the application of the Cepheids' pulsation period-luminosity relation from \citet{Wang2018}. 

We cross-match the luminosity distance data from \cite{Skowron2019a} with \textit{Gaia}'s DR3. This allows us to extract the proper motion and the three-dimensional (3D) velocity component of the Cepheids from the \textit{Gaia} dataset. However, we do not use the parallax information from \textit{Gaia} DR3. Additionally, we further limit the vertical position of the sample to $|z|<0.4$ \unit{kpc} using $z_{\odot} = 0.0208$ \unit{kpc}, and by restricting the sample to a maximum radial distance of 2 \unit{kpc} from the solar center. We specifically employ the same astrometry criterion as the one used for OB stars in the Cepheid dataset, which requires $\texttt{RUWE}<1.4$. The selection criteria results in a total of 154 Cepheid samples. 

\begin{figure}
	\includegraphics[width=\columnwidth ] {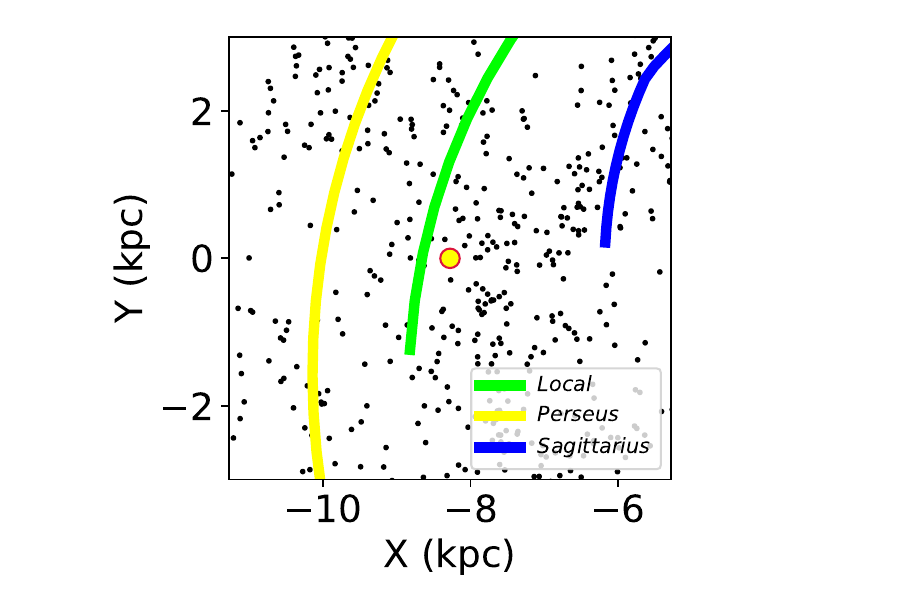}
    \caption{Galactocentric X-Y distribution of Cepheid overlaid with \citet{Reid2019} spiral arms. } 
    \label{XYdistribution(cepheids)}
\end{figure}

Fig. \ref{XYdistribution(cepheids)} depicts the distribution of Cepheids in the Galactocentric X-Y plane, with the spiral arms from \citet{Reid2019} superimposed. Due to the limited stars available, the Cepheids are more uniformly distributed without any noticeable concentration or bias towards particular regions as shown in Fig. \ref{XYdistribution(cepheids)}. Hence, we do not apply the homogenisation technique used for the OB stars in this case. 

Table \ref{tab:dataresults} presents our obtained value of $V_\mathrm{c}(R_0)$ as $239.16\pm2.52$ \unit{km.s^{-1}}. This value is similar to the results of OB stars within $\mathrm{D}<0.5$ kpc, $V_\mathrm{c}(R_0)= 238.30\pm2.24$ \unit{km.s^{-1}}. Furthermore, it is consistent with the result of $V_\mathrm{c}(R_0)= 236\pm3.0$ \unit{km.s^{-1}} reported by \citet{Kawata2019}, and \citet{Mroz2019} that obtained a value of $V_\mathrm{c}(R_0)= 233.6\pm2.8$ \unit{km.s^{-1}} for their Cepheid sample. We also obtain $V_{\phi,\odot}= 250.55\pm2.44$ \unit{km.s^{-1}}, which is consistent with the values found by our OB stars in $\mathrm{D}<0.5$ kpc and $\mathrm{D}<2$ kpc region.

The velocity value of $V_\mathrm{R,\odot}$ for Cepheid, which is $-8.77\pm1.20$ \unit{km.s^{-1}}, is slower than that of OB stars within both the $\mathrm{D}<0.5$ kpc and $\mathrm{D}<2$ kpc regions. However, this value falls within the uncertainty range of OB stars in the $\mathrm{D}<2$ kpc region and is consistent with the findings of \citet{Kawata2019}, who reported a $V_\mathrm{R,\odot}=-7.9\pm0.9$ \unit{km.s^{-1}} for Cepheid in the $\mathrm{D}<3$ kpc region.

The value of $\sigma_\mathrm{R}(R_0)$ for Cepheid is $14.34\pm0.85$ \unit{km.s^{-1}}. As discussed in the previous section, this value is lower compared to the value for OB stars in the $\mathrm{D}<2$ kpc region, $\sigma_\mathrm{R}(R_0) = 20.81\pm0.28$ \unit{km.s^{-1}}, but similar to the value of OB stars in the $\mathrm{D}<0.5$ kpc region, $\sigma_\mathrm{R}(R_0) = 14.83\pm0.27$ \unit{km.s^{-1}}. Overall, the value of $\sigma_\mathrm{R}(R_0)$ agrees with the value of $14.9\pm0.7$ \unit{km.s^{-1}} reported by \citet{Kawata2019}. The ratio $(\sigma_\phi/\sigma_\mathrm{R})^{2}$ is smaller for Cepheids compared to OB stars within the $\mathrm{D}<0.5$ kpc region, and significantly smaller compared to the value obtained within the $\mathrm{D}<2$ kpc region. The ratio $(\sigma_\phi/\sigma_\mathrm{R})^{2}$ is similar to what is found in \citet{Kawata2019}. As discussed in the previous section, this is much smaller than $(\sigma_\phi/\sigma_\mathrm{R})^{2}$ obtained from our OB stars sample within $\mathrm{D}<2$ \unit{kpc}.

We also obtain $R_{0}= 8.28\pm0.03$ \unit{kpc}, which is equally constrained for both Cepheid and OB stars by our strong prior of $R_{0}= 8.275\pm0.034$ \unit{kpc} \citep{GravityCollab2021}. In the Cepheid data, we find a negative $dV_\mathrm{c}(R_{0})/dR= -3.47\pm0.79$ \unit{km.s^{-1}.kpc^{-1}}, a value slightly more negative from what is obtained from our OB stars in the $\mathrm{D}<2$ kpc region, but consistent to the negative value achieved by \citet{Bobylev2017}, $dV_\mathrm{c}(R_{0})/dR= -3.6\pm1.7$ \unit{km.s^{-1}.kpc^{-1}}, and \citet{Kawata2019}, $dV_\mathrm{c}(R_{0})/dR= -3.7\pm0.5$ \unit{km.s^{-1}.kpc^{-1}}. This reassures that the overall trend of the circular velocity around the solar radius decreases with the radius in the scale of a few kpc.

\begin{figure}
	\includegraphics[width=\columnwidth]{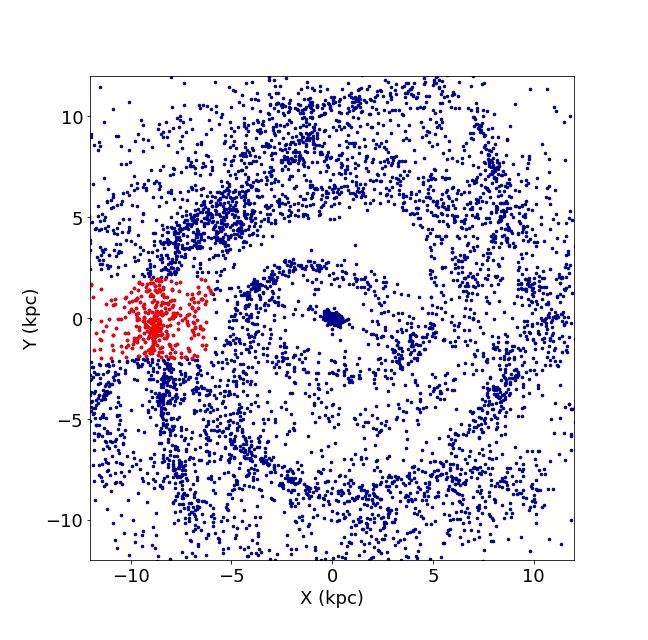}
    \caption{The blue points represent the Galactocentric X-Y distribution of star particles with ages $<$0.5~Gyr (blue dots) obtained from the \texttt{ASURA} simulation. The red points represent the selected data particles for analysis in this paper. } 
    \label{XY(simulation)}
\end{figure}

\subsection{Impacts of the spiral arm inferred from the N-body simulation data}

At the end of Section \ref{sec:Results}, we suggest that the observed positive $dV_\mathrm{c}(R_{0})/dR$ within 0.5 kpc from the Sun, in contrast to its negative $dV_\mathrm{c}(R_{0})/dR$ seen over a broader radial distance, can be attributed to the influence of the nearby spiral arm, specifically the Local arm. In this section, we compare the velocity structure of the young stars in a numerical simulation and compare it with our observational data. The simulation used is designed to model a disc structure similar to that of the Milky Way, using a numerical method that combines N-body and SPH, called \texttt{ASURA} \citep{Saitoh2008, Saitoh2009, Saitoh2010}. The simulation incorporates self-gravity, radiative cooling, star formation and stellar feedback as part of the modeling process, which is important for understanding the stellar kinematics and dynamical evolution of the disc galaxy. The simulation data used is from \citet{Baba2018} and the same snapshot as the one used in \citet{Kawata2019}. We have created our mock young star data from star particles whose ages are less than 0.5 Gyr. This stellar age includes stars with older ages than the OB stars. However, we select this larger range of age, to maximise the number of particles in our sample, without being influenced by excessively old stars. Fig. \ref{XY(simulation)} shows the face-on view of the distribution of the young stars in this simulated galaxy. We can see distinct spiral patterns traced by the young stars, as well as a prominent bar structure in the inner disk. However, the bar region displays limited ongoing star formation, except at its center and edges \citep[e.g.][]{Baba2022}. Consequently, the young stars predominantly represent the spiral arms rather than the entire bar structure.  In Fig. \ref{XY(simulation)}, the X and Y axes correspond to the galactocentric coordinates, where the galactic center is located at $(\textrm{X, Y})=(0, 0)$. The Y-axis is aligned with the direction of the observer's rotational velocity.  We select the young star particles around $(\mathrm{X, Y}) = (-8, 0)$ kpc, because there is a spiral arm just outside of this position. We choose the stars in the selected region of $-12<\mathrm{X}<-4$ kpc and $|\mathrm{Y}|<2$ kpc, which are highlighted with red dots in Fig. \ref{XY(simulation)}.

\begin{figure}
	\includegraphics[width=\columnwidth]{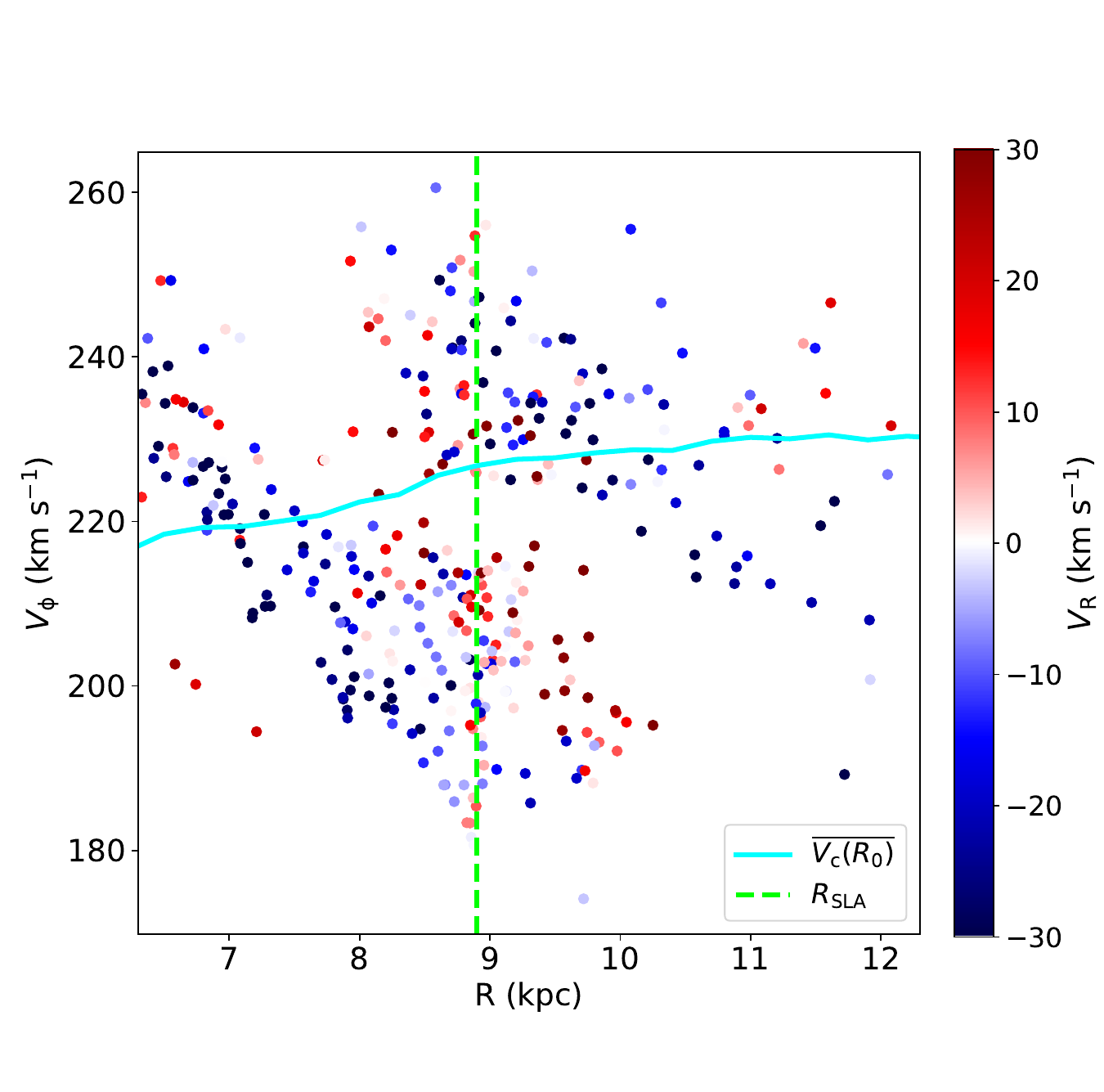}
    \caption{$R-V_{\mathrm{\phi}}$ distribution of the simulation star data colour-coded with $V_{\mathrm{R}}$. Cyan line represents the mean circular velocity, $\overline{V_\mathrm{c}(R_0)}$ of the simulation data. The vertical dashed green line represents the position of the spiral arm in the simulation, $R_{\mathrm{SLA}}$ = 8.9 \unit{kpc}. } 
    \label{r-vphi(simulation)}
\end{figure}

 Fig. \ref{r-vphi(simulation)} shows the rotation velocity of the selected young star particles as a function of the galactocentric radius colour coded with the radial velocity. From the number distribution of the selected particles as a function of radius, we find that there is a clear high density peak at $R_\mathrm{SLA}=8.9$ kpc, which we identify as a Local arm-like spiral structure in the simulated galaxy. We highlight the radius of this simulated Local arm-like spiral arm, $R_\mathrm{SLA}$, as the green dashed vertical line in Fig. \ref{r-vphi(simulation)}. In Fig. \ref{r-vphi(simulation)}, we see that inside the spiral arm, where $R<R_\mathrm{SLA}$, the stars are rotating slower and are moving in an outward direction (positive $\mathrm{V}_\mathrm{R}$), while outside of the spiral arm, the stars are rotating faster and are moving in an inward direction (negative $\mathrm{V}_\mathrm{R}$). This is a similar trend to what is observed in our OB stars data in Fig. \ref{r-vphi(OBstars)}. 

In Fig. \ref{r-vphi(OBstars)} the radius corresponding to the location of the Local arm is indicated with the vertical green dashed line. The radius of the Local arm is $R_{\mathrm{LA}}$ = 8.65 \unit{kpc}, where the locus of the Local arm in \citet{Reid2019} passes at Y=0 as shown in in Fig. \ref{XYdistribution}. The kinematic feature looks less clear in the observational data, perhaps due to data uncertainty and more data points. Also, the velocity difference around the arm is less in the observational data. For example, in Fig. \ref{r-vphi(OBstars)} the high rotation velocity of stars moving inward outside of the arm is around 250 \unit{km.s^{-1}}, and the low rotation velocity stars moving outward is 220 \unit{km.s^{-1}}, i.e. the difference is 30 \unit{km.s^{-1}}. On the other hand, for the simulation data shown in Fig. \ref{r-vphi(simulation)}, the high rotation velocity outside of the arm is around 240 \unit{km.s^{-1}}, and the low rotation velocity inside the arm is around 200 \unit{km.s^{-1}}, i.e. the difference is 40 \unit{km.s^{-1}}. The systematically lower rotation velocity of the simulation is caused by the lower mass used for the simulated galaxy than the Milky Way. The smaller difference in velocity around the arm in the observational data likely indicates that the Local arm is weaker than the arm in our N-body simulation. Yet, they display a comparable trend, suggesting that the Local arm resembles the spiral arm seen in the N-body simulations. The spiral arms in N-body simulations are always transient \citep[e.g.][]{Sellwood2011} and also co-rotating at all the radii \citep[e.g.][]{Wada2011, Grand2012a, DobbsandBaba2014}. Therefore, it may indicate that the Local arm could also be a transient spiral arm as seen in the N-body simulation. 

Interestingly, the N-body simulation shows a gap of stars between high and low rotation velocities around the spiral arm. The gap is probably due to the co-rotation resonance of the co-rotating spiral arm, suggesting that it may correspond to the circular velocity at $R=R_\mathrm{SLA}$. In our N-body simulation, we can measure the azimuthally mean circular velocity, $\overline{V_\mathrm{c}(R_0)}$, at different radii, because we can analyse the mass distribution of the whole galaxy. In contrast to the local $V_\mathrm{c}(R, \phi)$, the azimuthal mean $\overline{V_\mathrm{c}(R_0)}$ represents the total mass distribution of the Milky Way. Therefore, it is what we ultimately like to measure. In Fig. \ref{r-vphi(simulation)}, $\overline{V_\mathrm{c}(R_0)}$ is going through this gap feature between the outward slow-rotating stars and the inward fast-rotating stars at $\sim 8.9$  \unit{kpc}. 

\begin{figure}
	\includegraphics[width=\columnwidth]{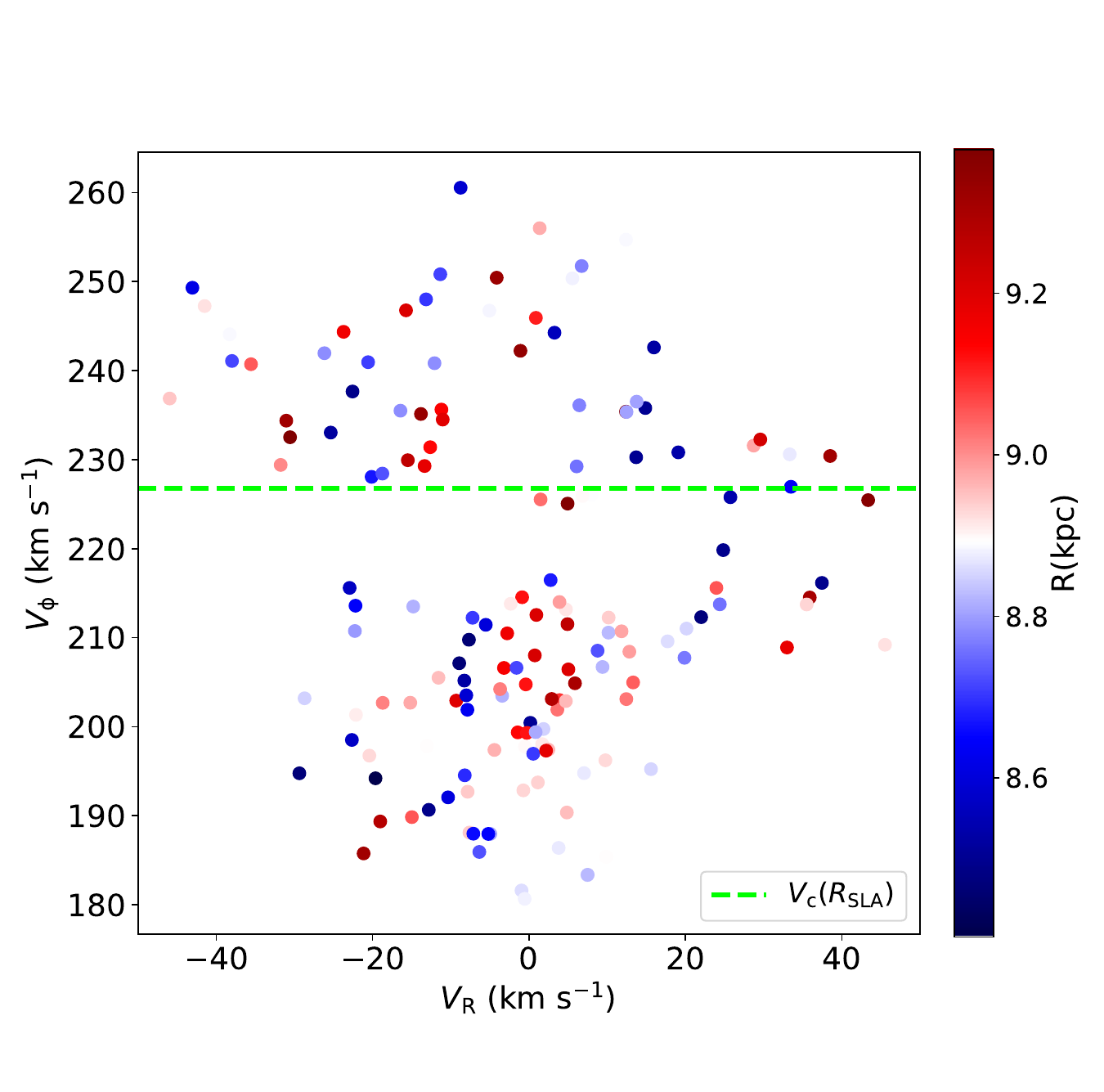}
    \caption{$V_{\mathrm{R}}-V_{\mathrm{\phi}}$ distribution of the simulation data colour-coded with $R$. The data selected here are only the stars close to the Local arm-like spiral structure, $[R_{\mathrm{SLA}}-0.5<R<R_{\mathrm{SLA}}+0.5]$, and with ages less than 0.5 Gyr. The green dashed line represents the the azimuthally averaged circular velocity at the radius corresponding to the Local arm-like spiral arm in the simulation, which is $V_\mathrm{c}(R_{\mathrm{SLA}})=226.767$  \unit{km.s^{-1}}.  } 
    \label{vr-vphi(simulation)}
\end{figure}

%Revisited in 8/8/2023
Furthermore, we closely look at the kinematics of the stars around the spiral arm in the simulation and plot the stars closer to the simulated Local arm-like spiral structure region, $[R_{\mathrm{SLA}}-0.5<R<R_{\mathrm{SLA}}+0.5]$. Fig. \ref{vr-vphi(simulation)} shows the rotation velocity, $V_\mathrm{\phi}$, against radial velocity, $V_\mathrm{R}$, for the selected star particles. The horizontal dashed line rotation velocity indicates the azimuthally mean $\overline{V_\mathrm{c}(R_{\mathrm{SLA}})}$. Between approximately $\sim220-225$ \unit{km s^{-1}}, there exists a noticeable gap in $V_{\mathrm{\phi}}$. The azimuthally mean circular velocity, $\overline{V_\mathrm{c}(R)}$, closely aligns with the upper boundary of this gap. This indicates that we can use this gap feature to infer the azimuthally mean circular velocity, although it only works at the location of the spiral arm.

\begin{figure}
	\includegraphics[width=\columnwidth]{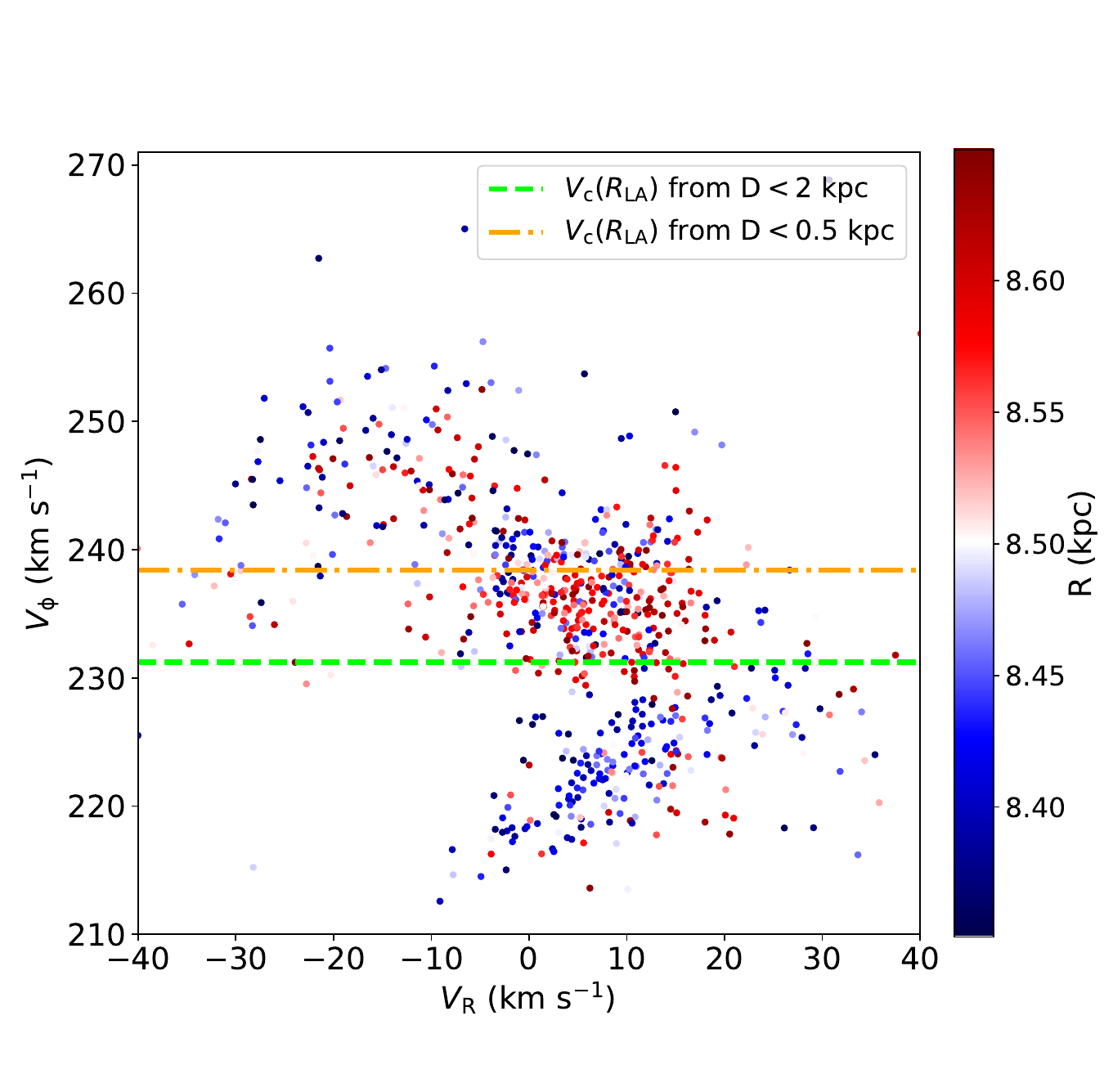}
    \caption{$V_{\mathrm{R}}-V_{\mathrm{\phi}}$ distribution of the OB stars data colour-coded with their respective $R$ distances. The data selected here are the stars close to the observed Local arm only, $[R_{\mathrm{LA}}-0.3<R<R_{\mathrm{LA}}+0.3]$ and $|Y|<0.2$~kpc. The orange dot-dashed line and green dashed lines represent the value of the Local arm's circular velocity, $V_\mathrm{c}R_{\mathrm{LA}}$ from $\mathrm{D}<0.5$ \unit{kpc} sample and $V_\mathrm{c}R_{\mathrm{LA}}$ from $\mathrm{D}<2$ \unit{kpc} sample.}
    \label{vr-vphi(OBstars)}
\end{figure}

Due to the resemblance between the observed Local arm and the simulated spiral arm, we examine the velocities $V_\mathrm{R}$ and $V_{\mathrm{\phi}}$ of the OB stars near the Local arm. Specifically, we focus on the stars within a narrow range of $[R_{\mathrm{LA}}-0.3<R<R_{\mathrm{LA}}+0.3]$ and $|\mathrm{Y}|<0.2$~kpc, where $R_{\mathrm{LA}}$ denotes the radius of the Local arm illustrated in Fig. \ref{vr-vphi(OBstars)}. Then, we see clear features of the known moving groups of Sirius ( $V_{\mathrm{\phi}}\sim240-260$  \unit{km.s^{-1}}), Coma Bernices ( $V_{\mathrm{\phi}}\sim230-240$ \unit{km.s^{-1}}) and Hyades-Pleiades ($V_{\mathrm{\phi}}\sim210-230$ \unit{km.s^{-1}}) in the OB stars. We find a clear gap (see Fig. \ref{vr-vphi(OBstars)}) between Coma Bernices and Hyades-Pleiades. This feature is also like what is seen around the spiral arm in the N-body simulation as shown in Fig. \ref{vr-vphi(simulation)}. Furthermore, the orange and green horizontal dashed lines in Fig. \ref{vr-vphi(OBstars)} show the $V_\mathrm{c}(R_\mathrm{LA})$ computed from our best-fit parameters in Table \ref{tab:dataresults} for $\mathrm{D}<0.5$ kpc and $\mathrm{D}<2$ kpc samples, respectively. Interestingly, $V_\mathrm{c}(R_\mathrm{LA})$ from $\mathrm{D}<2$ kpc best-fit parameter values are also well aligned with the upper boundary of the gap. 

Interestingly, Fig. \ref{vr-vphi(OBstars)} shows that the positive vertex deviation, i.e. the rotation velocity decreases with the radial velocity, in Coma Berenices moving group. On the other hand, the negative vertex deviation is seen in the Pleiades-Hyades moving group. However, the trend is much less clear and we need a further study with a higher resolution N-body simulations. 

In the N-body simulations, the spiral arms are co-rotating, and the gap is seen around the mean circular velocity, $\overline{V_\mathrm{c}(R_0)}$ of the galaxy because of the co-rotation resonance. The similarity of the location of the gap and $V_\mathrm{c}(R_\mathrm{LA})$ from $\mathrm{D}<2$ kpc sample to the trend seen in the simulation means that although $V_\mathrm{c}(R, \mathrm{\phi})$ is sensitive to the impact of the spiral arm, the gap of moving groups seen in the spiral arm is a reliable signature to identify the mean circular velocity, $\overline{V_\mathrm{c}(R_\mathrm{LA})}$. Then, our $V_\mathrm{c}(R_\mathrm{LA})$ from $\mathrm{D}<2$ kpc sample is likely to be close to the $\overline{V_\mathrm{c}(R_\mathrm{LA})}$ of the Galaxy. 

Conversely, the Sirius feature ($V_{\phi}\sim240-250$ \unit{km.s^{-1}}) is notably distant from the $\overline{V_\mathrm{c}(R_0)}$ in Fig. \ref{vr-vphi(OBstars)}. This feature is also seen as a strong resonance feature in the kinematics of the kinematically hot stars \citep{Kawata2021}. Hence, the Sirius feature is likely induced by the Galactic bar. It is interesting to see that the bar resonance also strongly impacts kinematics of the very young stars. This likely means that the orbit of the high-dense gas, i.e. star forming region, is already affected by the bar resonance. On the other hand, comparing the location of the strong resonance features in \citet{Kawata2021} and the known moving groups, we notice that there are no strong resonance features associated with Coma Bernice and Hyades-Pleiades. It could be because these two moving groups are induced by the Local arm rather than the resonance of the Galactic bar.

%%%%%%%%%%%%%%%%%%%%%%%%%
\section{Summary and Conclusions}
\label{sec:summary}
%%%%%%%%%%%%%%%%%%%%%%%%%

The Milky Way's circular velocity, specifically its value at the radius of the Sun, $V_\mathrm{c}(R_0)$, provides important insights into the Milky Way's dynamics and structure. However, measuring the value of $V_\mathrm{c}(R_0)$ is not straightforward, and different measurements yield systematically different results. $V_\mathrm{c}(R_0)$ is also affected by the presence of spiral arms. We measure $V_\mathrm{c}(R_0)$ by applying a simple axisymmetric model to the kinematics of the OB stars in the different sizes of the regions and assess the impact of the Local arm. 

We fit the kinematics of the OB stars measured by \textit{Gaia} \citep{Drimmel2022gaia} with the 2D axisymmetric model following \citet{Kawata2019} using MCMC. We take into account the observational uncertainty and asymmetric drift. This allows us to measure $V_\mathrm{c}(R_0)$ at the solar radius, $R_0$, the radial velocity, $V_\mathrm{R,\odot}$, and the azimuthal velocity, $V_{\phi,\odot}$, of the Sun and the slope of the circular velocity, $dV_\mathrm{c}(R_0)/dR$. In our model, we use strong priors, $R_{0, \mathrm{prior}} = 8.275\pm0.003$ \unit{kpc} from \citet{GravityCollab2021}, and the Sun’s angular rotation velocity, $\Omega_{\odot} = 30.32 \pm 0.27$ \unit{km.s^{-1}.kpc^{-1}} from \citet{ReidBrunthaler2020}. We fit the OB stars data within $\mathrm{D}<0.5$ \unit{kpc} and $\mathrm{D}<2$ \unit{kpc} from the Sun separately. 

For $\mathrm{D}<2$ \unit{kpc}, we have obtained $V_\mathrm{c}(R_0)= 233.95\pm2.24$ \unit{km.s^{-1}}, $V_{\phi,\odot}= 247.74\pm2.22$ \unit{km.s^{-1}}, $V_\mathrm{R,\odot}=-9.65\pm0.43$ \unit{km.s^{-1}}, and $R_0=8.28\pm0.04$ \unit{kpc}. We compare this result to previous studies that use a different set of tracer population stars. In general, our results are consistent with the recent studies (see Fig. \ref{resultsVsliterature} for a summary). 

Since using different tracer population yield different results, we further benchmark our results with Cepheids data by cross-matching the luminosity distance data from \cite{Skowron2019a} with \textit{Gaia}'s DR3. Using the same MCMC analysis and priors, we find that the overall results of our OB stars fall within the uncertainty range of Cepheids. Our best-fit results are also consistent with that of \citet{Kawata2019}. This reassures that the overall trend of the circular velocity around the solar radius decreases with the radius in the scale of a few kpcs.  Moreover, our best fit result of $V_\mathrm{c}(R_0)$ with a negative $dV_\mathrm{c}(R_0)/dR$ slope in $\mathrm{D}<2$ \unit{kpc} is also consistent with the overall negative trend of median $V_{\phi, \mathrm {R}}$ as shown in Fig. \ref{r-vphi(OBstars)}.

However, in $\mathrm{D}<0.5$ \unit{kpc}, our best fit result shows a positive  $dV_\mathrm{c}(R_0)/dR$ slope, which is consistent with the median $V_{\phi,\mathrm{R}}$ in the $[-0.5+R_{0}<R<R_{0}+0.5]$ kpc region. This also drives a slightly higher $V_\mathrm{c}(R_0)$ than the best-fit value for the data within $\mathrm{D}<2$ \unit{kpc}. In this radial range, $V_\mathrm{R}$ changes from a positive value (outward motion) to a negative value (inward motion) as the radius increases. This is a typical velocity trend seen around the spiral arm in the numerical simulations \citep[e.g.][]{Grand2012b, Baba2013} whose spiral arms are co-rotating, dynamic and transient. Hence, we compare the velocity structure of young stars obtained from the N-body/SPH numerical simulation to our OB stars data and find a resemblance of stellar kinematics around the observed Local arm and the simulated spiral arm by delving deeper into the velocities, $V_\mathrm{R}$ and $V_{\phi}$ of the OB stars in proximity to the spiral arm. 

We further analyse $V_\mathrm{R}$ and $V_{\phi}$ distribution of OB stars within the radial range of $[R_{\mathrm{LA}}-0.3 < R < R_{\mathrm{LA}}+0.3]$, with $R_{\mathrm{LA}}$ signifying the Local arm's radius. Within this radial region, prominent traits of the recognized moving groups, namely Sirius, Coma Bernices, and Hyades-Pleiades, are observed. We also find the known gap of $V_{\phi}$ between Coma-Bernices and Hyades-Pleiades moving groups. Similarly, we analyse $V_\mathrm{R}$ and $V_{\phi}$ distributions of the young stars around the spiral arm in our simulation. This analysis reveals a comparable gap in the distribution of $V_{\phi}$. As per the simulation, spiral arms co-rotate, and the gap is attributed to the co-rotation resonance, and therefore this gap is around the Galaxy's mean circular velocity, $\overline{V_\mathrm{c}(R_0)}$, at the spiral arm. This suggests that the observed gap in moving groups within the arm serves as a dependable marker for discerning the mean circular velocity, $\overline{V_\mathrm{c}(R_\mathrm{SLA})}$. Consequently, our derived $V_\mathrm{c}(R_\mathrm{LA})$ from the $\mathrm{D} < 2$ kpc data is found to be aligned well at the gap at the Local arm, and this likely indicates that the circular velocity derived from $\mathrm{D}<2$ \unit{kpc} sample is closer to the azimuthally averaged circular velocity, because it is caused by the co-rotation resonance of the Local arm at the radius. 

Our research suggests that the moving groups, namely Coma Bernices and Hyades-Pleiades, are likely induced by the presence of the Local arm. As a result, the Local arm is more of a major spiral arm than a branch-like feature or a minor spur only traced by gas and very young stars. A more in-depth analysis can be undertaken by assessing the stellar density enhancement of stars within proximity to the Local arm. \citet{Miyachi2019} pioneered the assessment of the stellar density excess around the Local arm using \textit{Gaia} data, carefully taking into account the completeness of the data. However, \citet{Miyachi2019} constrain their study within $90^\circ<l<270^\circ$ in order to avoid the complexity of high extinction. Furthermore, they restrict their data analysis to $\mathrm{D}<1.3$ \unit{kpc} due to the completeness constraints of the \textit{Gaia} dataset. We need a more advanced statistical model to take into account the observational selection function to trace the Local arm in a larger radial range. Revealing the stellar density structure of the Local arm in the larger radial region should enable us to determine if the Local arm indeed represents a major spiral arm. Furthermore, such a study should be coupled with stellar velocity distribution around the Local arm \citep[e.g.][]{Liu2017} over a larger radius range. This insight could clarify the character of the Local arm, revealing whether it resembles a density wave or has traits of a dynamic spiral arm.

\section*{Acknowledgements}
ASA acknowledges the funding body, the UAE Ministry of Presidential Affairs, for their support through the PhD scholarship. This work was partly supported by the UK's Science \& Technology Facilities Council (STFC grant ST/S000216/1, ST/W001136/1).
This work is a part of MWGaiaDN, a Horizon Europe Marie Sk\l{}odowska-Curie Actions Doctoral Network funded under grant agreement no. 101072454 and also funded by UK Research and Innovation (EP/X031756/1). This work has made use of data from the European Space Agency (ESA) mission {\it Gaia} (\url{https://www.cosmos.esa.int/gaia}), processed by the {\it Gaia}
Data Processing and Analysis Consortium (DPAC, \url{https://www.cosmos.esa.int/web/gaia/dpac/consortium}). Funding for the DPAC has been provided by national institutions, in particular the institutions participating in the {\it Gaia} Multilateral Agreement.

%%%%%%%%%%%%%%%%%%%%%%%%%%%%%%%%%%%%%%%%%%%%%%%%%%
\section*{Data Availability}

The pre-processing data and code scripts used in this work can be made available upon request.

%The inclusion of a Data Availability Statement is a requirement for articles published in MNRAS. Data Availability Statements provide a standardised format for readers to understand the availability of data underlying the research results described in the article. The statement may refer to original data generated in the course of the study or to third-party data analysed in the article. The statement should describe and provide means of access, where possible, by linking to the data or providing the required accession numbers for the relevant databases or DOIs.



\bibliographystyle{mnras}
\bibliography{bibliography} % if your bibtex file is called example.bib

%\begin{thebibliography}{99}
%\end{thebibliography}

%%%%%%%%%%%%%%%%%%%%%%%%%%%%%%%%%%%%%%%%%%%%%%%%%%

%%%%%%%%%%%%%%%%% APPENDICES %%%%%%%%%%%%%%%%%%%%%

%\appendix



%%%%%%%%%%%%%%%%%%%%%%%%%%%%%%%%%%%%%%%%%%%%%%%%%%

% Don't change these lines
\bsp	% typesetting comment
\label{lastpage}
\end{document}